\documentclass[aps,prb,twocolumn,superscriptaddress,floatfix,longbibliography]{revtex4-2}

\usepackage[utf8]{inputenc}
\usepackage[T1]{fontenc}
\usepackage{amsmath, amssymb, bm}
\usepackage{graphicx}
\usepackage{xcolor}
\usepackage{hyperref}
\usepackage{cleveref}
\usepackage{booktabs}
\usepackage{algorithmic}
\usepackage{physics}
\usepackage{siunitx}

\hypersetup{
    colorlinks,
    linkcolor={blue!50!black},
    citecolor={blue!50!black},
    urlcolor={blue!80!black},
}

\newcommand{\vR}{\mathbf{R}}
\renewcommand{\vr}{\mathbf{r}}
\newcommand{\vk}{\mathbf{k}}
\newcommand{\vh}{\mathbf{h}}

\newcommand{\MLP}{\operatorname{MLP}}
\newcommand{\RMSNorm}{\operatorname{RMSNorm}}
\newcommand{\EdgeAttn}{\operatorname*{EdgeAttn}}

\begin{document}

\title{Continuum Neural Momentum Eigenstate for Variationally Solving Quasiparticles}

\author{David D. Dai}
\email{dddai@mit.edu}
\author{Marin Solja\v{c}i\'c}
\email{soljacic@mit.edu}
\affiliation{Department of Physics, Massachusetts Institute of Technology, Cambridge, Massachusetts 02139, USA}
\affiliation{The NSF AI Institute for Artificial Intelligence and Fundamental Interactions}

\date{\today}

\begin{abstract}
We design the first neural quantum state for continuum particles that, for any chosen allowed momentum $\vk$, is by construction an exact eigenstate of total momentum with eigenvalue $\vk$.
Our architecture, EVE, enables off-the-shelf VMC to solve for momentum-sector ground states.
We test EVE on 2D bosons with mutual $1/r$ interactions, finding that a single unified ansatz is capable of describing four qualitatively different states: superfluid, roton, crystal, and phonon.
At different densities, we extract the underlying phase of matter from the dispersion's shape.
At $r_s = 20.0$, we see the roton minimum at finite $k$ expected of a superfluid.
At $r_s = 100.0$, we see striking zone folding indicative of crystalline order, with periodically spaced minima representing floating crystals connected by phonon arcs in between.
Using density-density correlation functions, we confirm the phase diagnoses and probe the excitations' correlation structures.
Finally, we analyze the roton's phase texture and find unexpected multi-particle phase strings, formed when several vortex dipoles merge, leaving two vortices connected by a phase slip.
\end{abstract}

\maketitle

\section{Introduction}
    Understanding the energy-momentum relationship of a system's quasiparticles is a core pillar of condensed matter physics.
    From Landau, Feynman, and Cohen's work on phonons and rotons in superfluid $^4\mathrm{He}$~\cite{Landau_1941_superfluidity,Feynman_1954_theory,Feynman_1956_backflow}, to Nambu and Goldstone's bosons in symmetry-breaking phases~\cite{Nambu_1960_quasiparticle,Goldstone_1961_field_theory,Goldstone_1962_broken}, to the modern discovery of fractionalization~\cite{Tsui_1982_experiment,Laughlin_1983_theory,Arovas_1984_statistics,Moore_1991_nonabelian} and topological order~\cite{Wen_1990_order,Wen_1990_degeneracy}, understanding the rich and often unexpected behavior of low-lying excitations has driven much progress in our field.
    However, computationally modeling quasiparticles is difficult.

    Variational Monte Carlo (VMC)~\cite{Feynman_1954_theory,McMillan_1965_ground,Ceperley_1977_first,Ceperley_1980_electron,Tanatar_1989_2DEG,Foulkes_2001_review,Needs_2010_review} with a generic ansatz yields the global ground state, so excited-state methods~\cite{Zhao_2016_shift,Choo_2018_orthogonalize,Pathak_2021_penalty,Entwistle_2023_penalty,Wheeler_2024_ensemble,Pfau_2024_excited} must constantly suppress contamination by lower-lying states, such as by adding an overlap penalty to the loss~\cite{Pathak_2021_penalty,Entwistle_2023_penalty,Wheeler_2024_ensemble}.
    Such methods add hyperparameters, are more complex, and are less reliable than standard VMC.
    Furthermore, for finding momentum-sector ground states, which in most cases correspond to the lowest quasiparticle branch, even a perfect general-excited-state method is not ideal.
    Due to irrelevant vertical excitations and point-group degeneracy among symmetry-related momenta, only a small fraction of naive excited states actually yield new points on the dispersion curve.
    On the other hand, specialized target-momentum-sector methods exist~\cite{Lange_2024_quasiparticles,Choo_2018_orthogonalize,Romero_2025_spectroscopy,Yoshioka_2021_band,Fu_2024_qiankun,Zhang_2025_NTB} but rely on adding quantum-number penalties to the loss~\cite{Lange_2024_quasiparticles}, only apply to discrete lattice models~\cite{Choo_2018_orthogonalize,Romero_2025_spectroscopy}, or work in a truncated second quantization basis~\cite{Yoshioka_2021_band,Fu_2024_qiankun,Zhang_2025_NTB}.

    Here, we introduce EVE, a neural quantum state (NQS)~\cite{Carleo_2017_solving,Nomura_2017_RBM,Carleo_2019_NetKet,Choo_2019_CNN,Sharir_2020_RNN,Pfau_2020_FermiNet,Hermann_2020_PauliNet,Spencer_2020_better,Viteritti_2023_spinformer,VonGlehn_2023_PsiFormer,Hermann_2023_review,Lange_2024_review} for continuum bosons that is an exact momentum eigenstate with configurable eigenvalue.
    Architecturally, EVE ingests pairwise features, which are naturally translation invariant, and transforms them using efficient $\order{N^2}$-scaling edge-to-vertex and vertex-to-edge (thus the name EVE, short for edge-vertex-edge) updates into rich one- and two-particle representations.
    The readout step then imparts the desired momentum using a phase factor.
    EVE integrates directly into off-the-shelf VMC without adding hyperparameters, and it sidesteps basis-set truncation error by working directly in the continuum.
    EVE brings to bear the full power of NQS---which has been so successful for the ground-state problem---onto the quasiparticle problem.
    
    For the 2D boson gas with $1/r$ interactions, we find quantitatively accurate results for both ground and excited states of the superfluid and Wigner crystal using a single EVE model without phase-specific biases.
    For the superfluid, we see the roton minimum and directly study the roton's phase texture, finding both the expected phase dipoles at particle coalescences but also unexpected multiparticle phase-string structures.
    For the crystal, we see an emergent Brillouin zone and zone folding, with an oscillating dispersion that has minima at the expected Anderson tower sites.
    In all situations, we estimate from zero-variance extrapolation~\cite{Fu_2024_variance} that the relative error in the \textit{dispersion} is $1\%$ or less, with much lower relative error in the \textit{total energy}.

\section{Neural Momentum Eigenstate}

    We study $N$ bosons in a 2D periodic box with direct lattice vectors $\mathbf{a}_i$.
    The bosons experience mutual $1/r$ repulsion and carry no internal degrees of freedom.
    The Hamiltonian is
    \begin{equation}
        H = -\frac{1}{2r_s} \sum_i \nabla^2_i + \sum_{i < j} \frac{1}{|\vr_i - \vr_j|},
    \end{equation}
    where $r_s$ is the density parameter, and larger values mean stronger interactions~\footnote{To stabilize the overall energy scale in the low-density regime, we choose $\mathbf{a}_i$ such that the area per particle is always $\pi$. In other words, we scan the phase diagram by changing the kinetic energy's coefficient instead of scaling the system area.}.
    By translation invariance, we can choose energy eigenstates with definite momenta $\mathbf{k}$.

    \subsection{Architecture}

\begin{figure}[t!]
    \centering
    \includegraphics[height=0.4\textwidth]{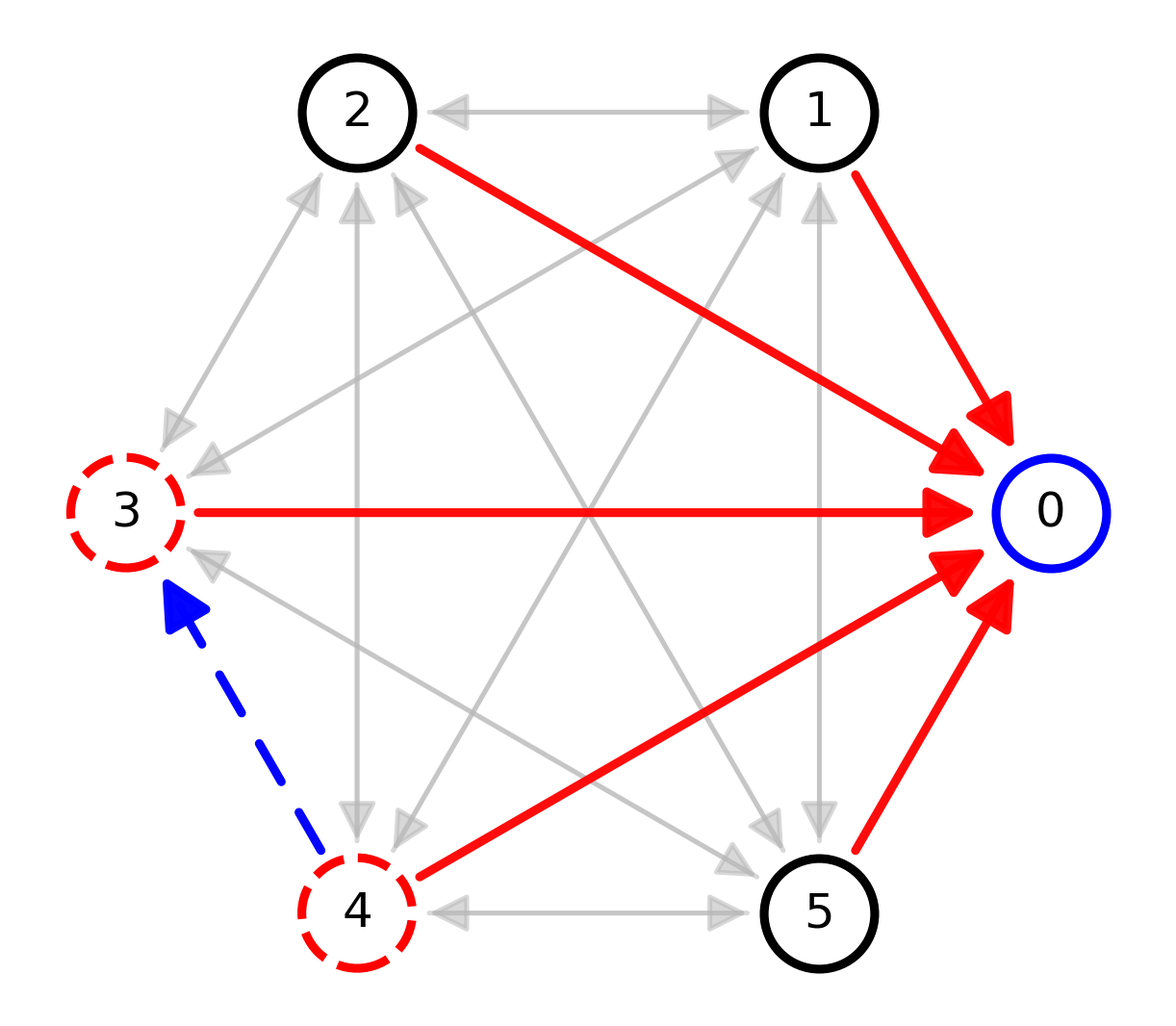}
    \caption{Solid lines show the edge-to-vertex aggregation update, which pools edges $\vh_{01}$ to $\vh_{05}$ (solid red arrows) into vertex $\vh_0$ (solid blue circle). Dashed lines show the vertex-to-edge feedback, where vertices $\vh_3$ and $\vh_4$ (dashed red circles) modify edge $\vh_{34}$ (dashed blue arrow).}
    \label{fig:graph}
\end{figure}

        EVE's high-level form for momentum sector $\vk$ is:
        \begin{equation}\label{eq:EVE_head}
            \begin{aligned}
                \Psi\left(\{\vr\}\right) =& 
                \sum_{i} e^{i \mathbf{k} \cdot \vr_i} f\left(\vr_i; \{\vr_{/i}\}\right)\\
                &\times e^{J\left(\{\vr\}\right)}
                \prod_{i} g\left(\vr_i; \{\vr_{/i}\}\right).
            \end{aligned}
        \end{equation}
        Throughout, $\{\vr\}$ denotes the set of all particle coordinates, and $\{\vr_{/i}\}$ denotes the same but excluding $i$.
        $J$ is a permutation and translation-invariant Jastrow:
        \begin{equation}\label{eq:jastrow_conditions}
            \begin{aligned}
                J\left(\{\vr\}\right) &\equiv J\left(\vr_1, \vr_2 \ldots \vr_N\right)\\ 
                &= J\left(\vr_{\sigma_1}, \vr_{\sigma_2} \ldots \vr_{\sigma_N}\right),\\
                J\left(\{\vr + \Delta \vr\}\right) &= J\left(\{\vr\}\right).
            \end{aligned}
        \end{equation}
        $f$ and $g$ are translation-invariant generalized orbitals:
        \begin{equation}\label{eq:generalized_orbital}
            \begin{aligned}
                f\left(\vr_1; \{\vr_{/1}\}\right) \equiv & f\left(\vr_1;\vr_2 \ldots \vr_N\right) \\
                =& f\left(\vr_1;\vr_{\sigma_2} \ldots \vr_{\sigma_N}\right),\\
                f\left(\vr_1 + \Delta \vr; \{\vr_{/1} + \Delta \vr\}\right) &= f\left(\vr_1; \{\vr_{/1}\}\right).\\
            \end{aligned}
        \end{equation}
        In Eqs. \ref{eq:jastrow_conditions} and \ref{eq:generalized_orbital}, $\sigma$ is a permutation of particle labels and $\Delta \vr$ is a displacement applied to all particles.
        Eq. \ref{eq:EVE_head} is universal and recovers an arbitrary momentum-$\vk$ $\Psi$ as
        \begin{equation}
            \Psi(\{\vr\}) = \sum_{i} e^{i \mathbf{k} \cdot \vr_i} \left(\frac{e^{-i \mathbf{k} \cdot \vr_i} \Psi(\{\vr\})}{N}\right).
        \end{equation}
        Because $J$, $f$, and $g$ are translation-invariant functions, we can construct them using only pairwise $\vr_i - \vr_j$.
        
\begin{figure}[t!]
    \centering
    \includegraphics[height=0.4\textwidth]{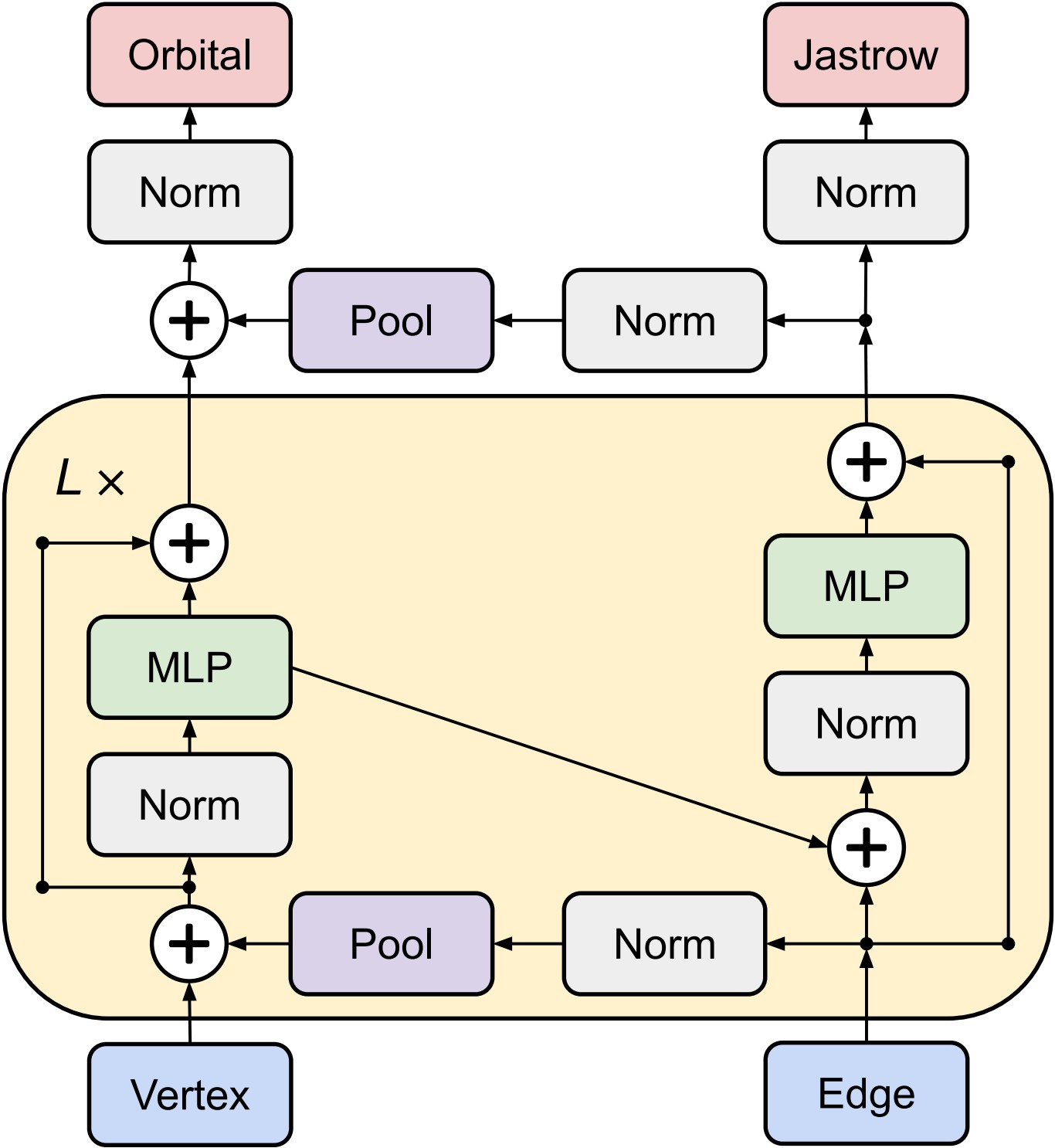}
    \caption{High-level schematic of the EVE architecture. $\operatorname{Pool}$ denotes $\EdgeAttn$ and $\operatorname{Norm}$ denotes $\RMSNorm$. Each block transforms input vertex and edge features through edge-to-vertex pooling, in-place vertex update, and vertex-perturbed in-place edge update.}
    \label{fig:block}
\end{figure}
        
        EVE operates on a fully connected graph with vertices $\vh_i$ and directed edges $\vh_{ij}$.
        Our first update is an attention-like~\cite{Bahdanau_2014_align,Luong_2015_attention,Vaswani_2017_transformer} edge-to-vertex learned pooling:
        \begin{equation}\label{eq:EdgeAttn}
            \begin{gathered}
                w_{ij}^h = \frac{e^{z_{ij}^h}}{\sum_{k \neq i} e^{z_{ik}^h}}, \quad z_{ij}^h = \mathbf{b}_\text{L}^h \cdot \vh_{ij}, \quad \mathbf{v}_{ij}^h = \mathbf{W}^h_\mathrm{V} \vh_{ij},\\
                \EdgeAttn_j(\vh_{ij}) = \mathbf{W}_\mathrm{O} \operatorname*{concat}_h \left[ \sum_{j \neq i} w_{ij}^h \mathbf{v}_{ij}^h \right],\\
            \end{gathered}
        \end{equation}
        where $h$ indexes among several attention heads. Compared to traditional self-attention, we form logits directly from edge features instead of a $QK$ dot product.
        
        For in-place updates, we use feedforward $\MLP(\vh) =\mathbf{W}_2 \operatorname{GELU}(\mathbf{W}_1 \vh)$, where $\operatorname{GELU}$ is an element-wise activation~\cite{Hendrycks_2016_GELU}.
        For normalization, we use $\RMSNorm(\vh) = \vh/\sqrt{\epsilon + \norm{\vh}_2^2}$ prior to sublayers~\cite{Zhang_2019_RMSNorm,Xiong_2020_prenorm}.
        In all cases, $\operatorname{MLP}$ and $\operatorname{RMSNorm}$ act individually on and do not mix between different vertices and edges.   

\begin{figure*}[t!]
    \centering
    \begin{minipage}{0.495\textwidth}
        \centering
        \includegraphics[height=0.8\linewidth]{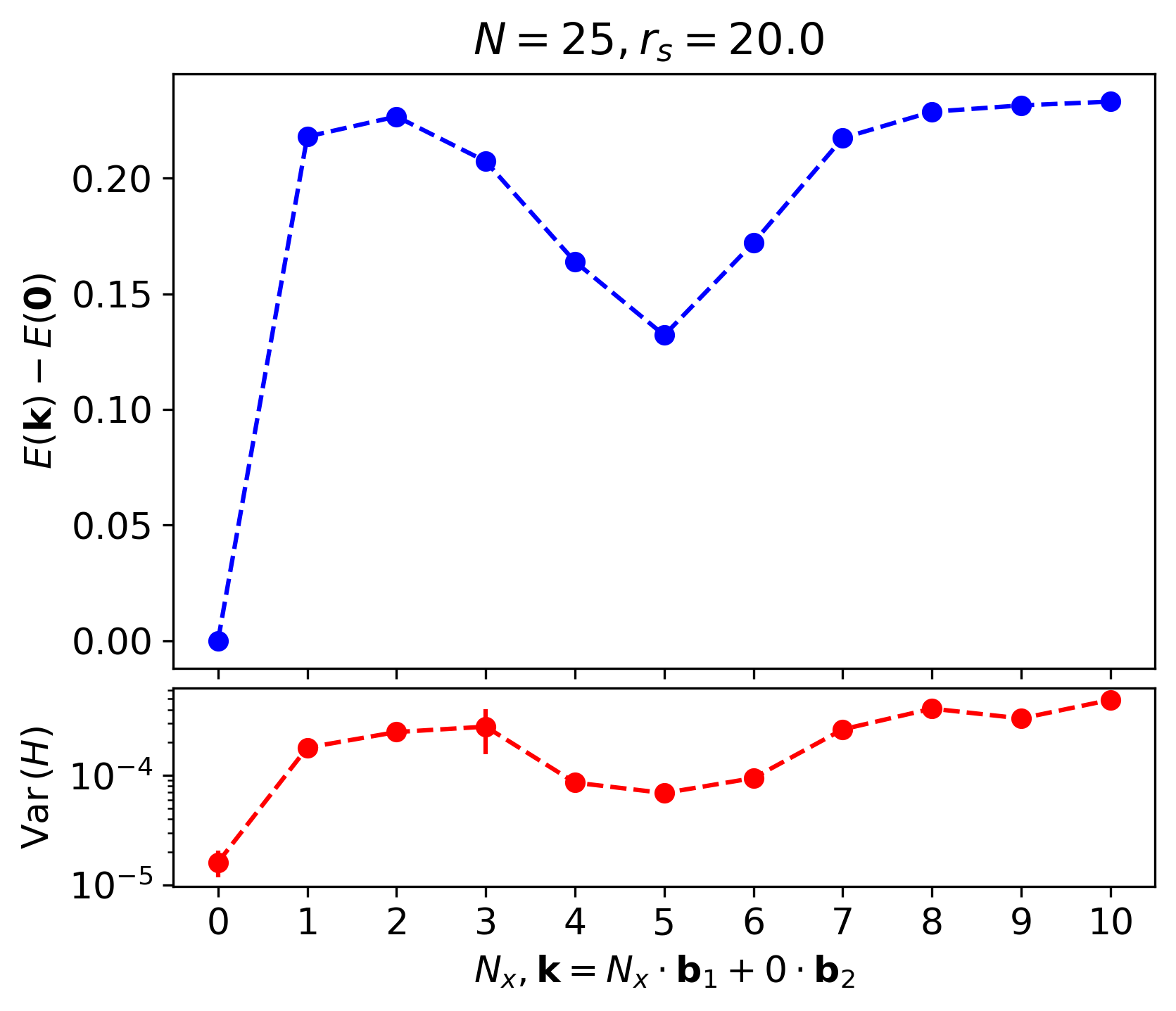}
    \end{minipage}
    \hfill
    \begin{minipage}{0.495\textwidth}
        \centering
        \includegraphics[height=0.8\linewidth]{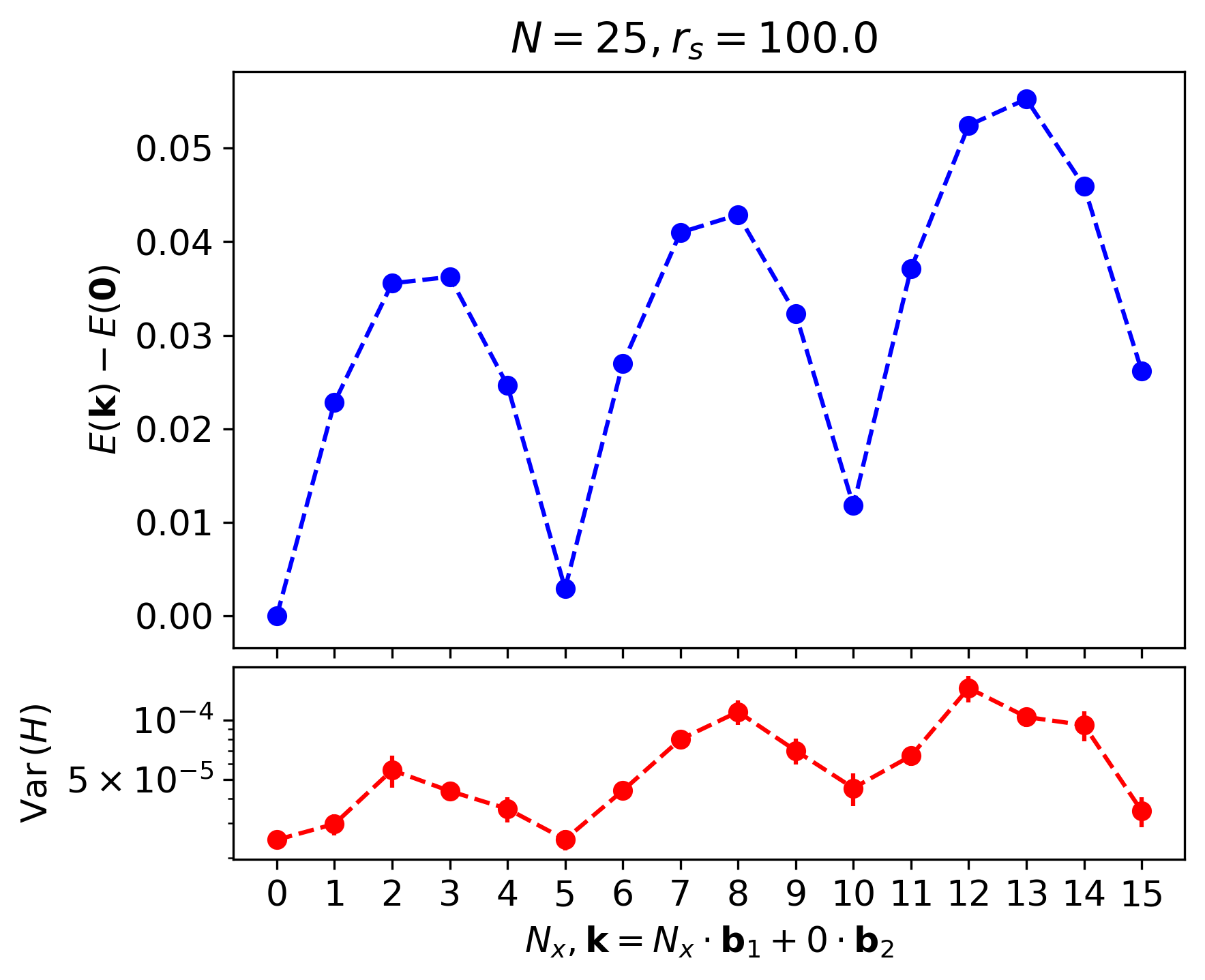}
    \end{minipage}
    \begin{minipage}{0.495\textwidth}
        \centering
        \includegraphics[height=0.635\linewidth]{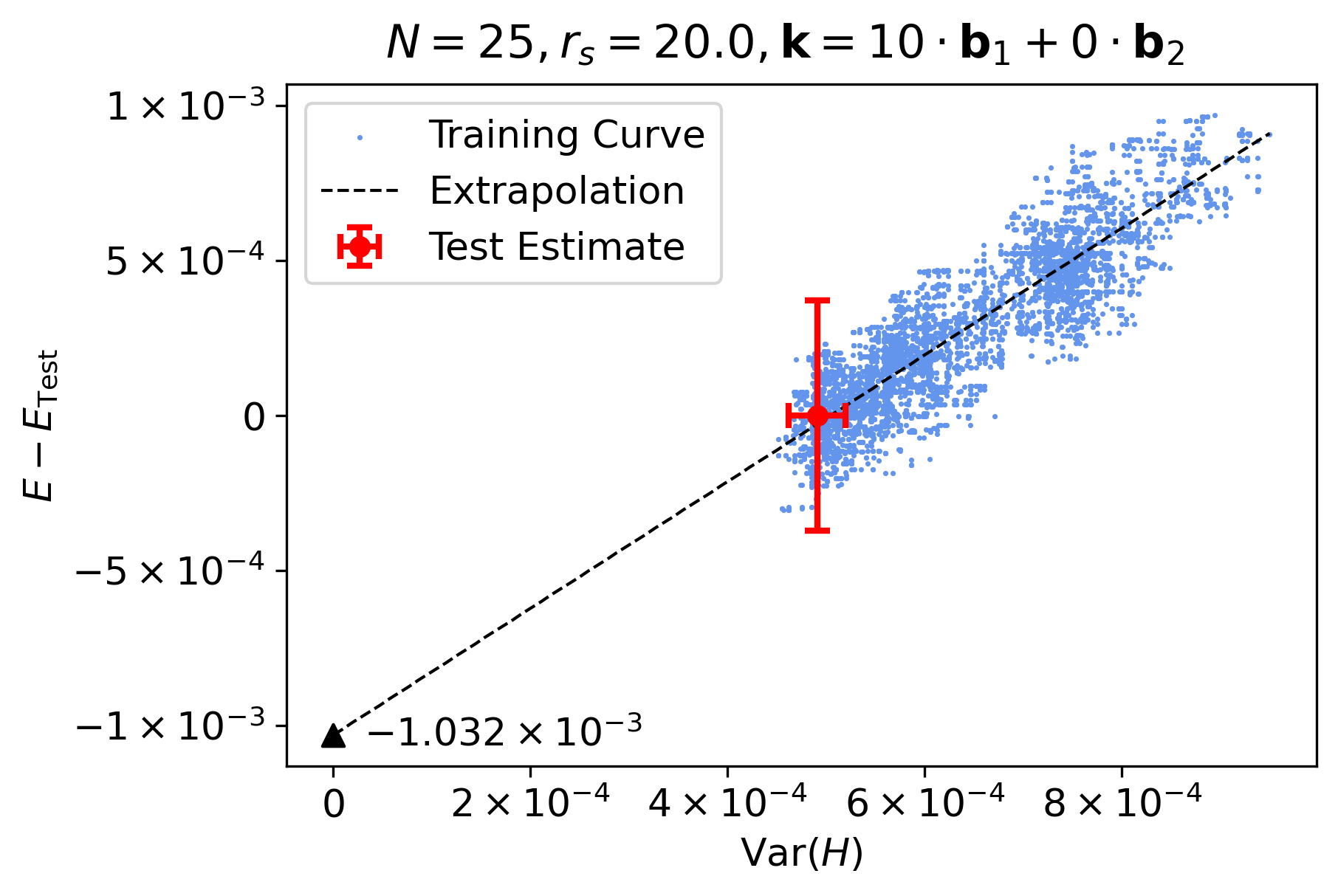}
    \end{minipage}
    \hfill
    \begin{minipage}{0.495\textwidth}
        \centering
        \includegraphics[height=0.635\linewidth]{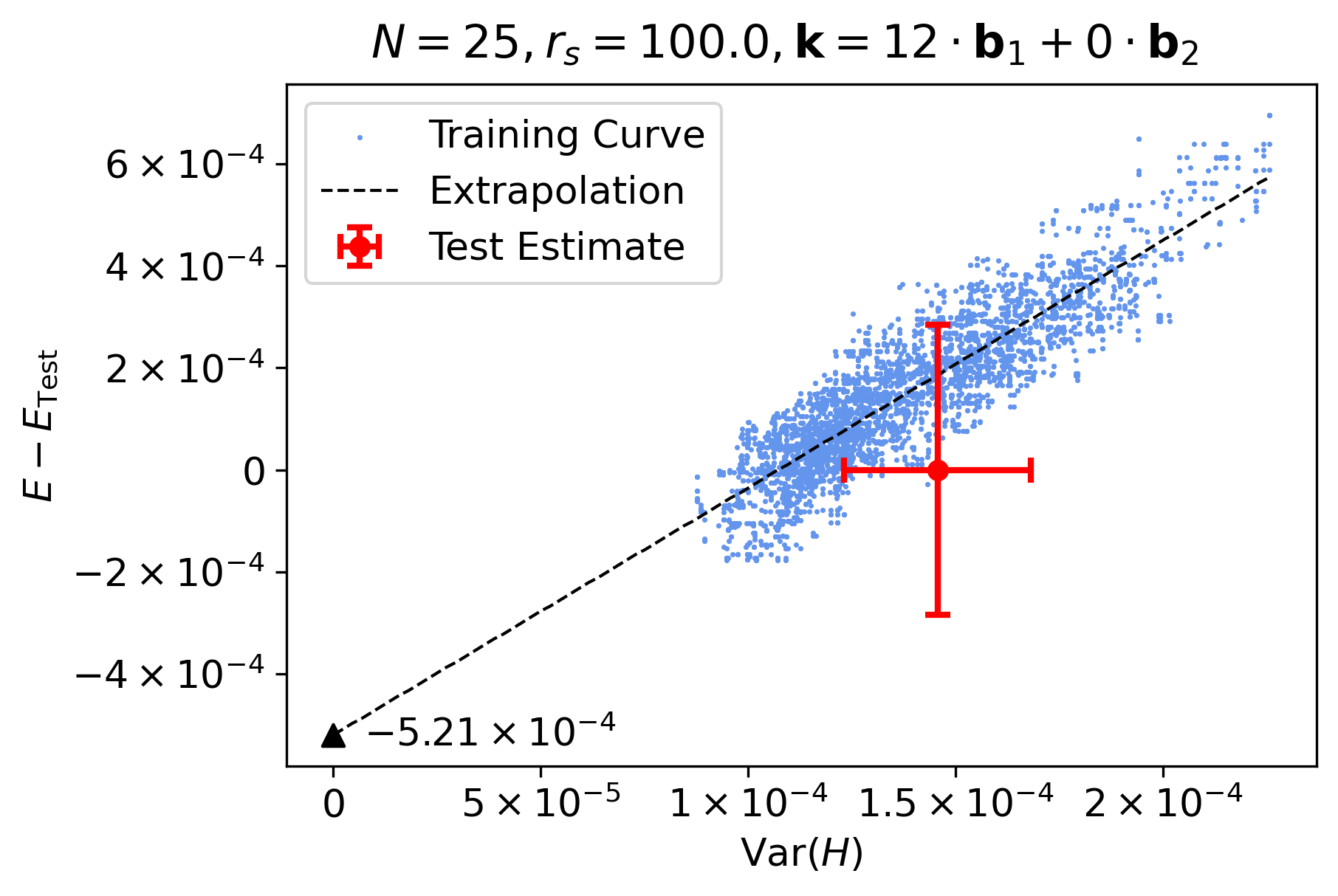}
    \end{minipage}
    \caption{%
    Top Left: The superfluid's momentum-sector energies. A pronounced roton minimum exists near $N_x = 5$.
    Top Right: The Wigner crystal's momentum-sector energies. The minima at $N_x = 0,5,10,15$ correspond to Anderson tower states, in this case floating crystals with different center-of-mass momenta, while the arcs are phonon excitations.
    Bottom Left: Error estimation for the superfluid's highest-variance sector.
    Bottom Right: Error estimation for the Wigner crystal's highest-variance sector.
    We infer dispersions after training using freshly burned-in walkers that the model has never seen before.
    For both extrapolations, the scattered points are training measurements smoothed by taking the median over a $25$-step window.}
    \label{fig:energy}
\end{figure*}
        
        A full EVE block performs
        \begin{equation}
            \begin{gathered}
                \vh_i \leftarrow \vh_i + \EdgeAttn_j(\RMSNorm(\vh_{ij})),\\
                \mathbf{y}_k, \mathbf{y}^\mathrm{row}_k, \mathbf{y}^\mathrm{col}_k = \MLP_1(\RMSNorm{\left(\mathbf{h}_k\right)}),\\
                \vh_i \leftarrow \vh_i + \mathbf{y}_i,\\
                \vh_{ij} \leftarrow \vh_{ij} + \MLP_2(\RMSNorm(\vh_{ij} + \mathbf{y}^\mathrm{row}_i + \mathbf{y}^\mathrm{col}_j)).\\
            \end{gathered}
        \end{equation}
        In words, EVE first aggregates edges into vertices.
        Then, it performs a vertex MLP that both prepares residual updates for the vertex stream and row and column updates for the edge stream.
        Finally, the row and column updates perturb the edge MLP's input, whose output updates the edge stream.
        Fig. \ref{fig:graph} shows the two graph interactions, and Fig. \ref{fig:block} shows the overall information flow in the EVE architecture.
        Our approach globally contextualizes the vertex and edge features at $\order{N^2}$ cost.

        The input vertex features are $\vh^0_{i} = \mathbf{0}$ to preserve translation invariance.
        Define torus angles $\vr = \sum_{\mu=1}^d \frac{\phi_\mu}{2\pi} \mathbf{a}_\mu$ and metric $S_{\mu\nu} = \mathbf{a}_\mu \cdot \mathbf{a}_\nu$. The input edge features are
        \begin{equation}
            \begin{aligned}
                s(r) &= \log\left(1 - \alpha_\mathrm{soft} + \sqrt{\alpha_\mathrm{soft}^2 + r^2}\right),\\
                \vh^0_{ij} &= \mathbf{W}_\mathrm{E} \operatorname*{concat}_\mu \left[s\left(\norm{\vr}_{\mathbf{a}}\right), (\cos{\phi_\mu}, \sin{\phi_\mu}) \right].\\
            \end{aligned}
        \end{equation}
        We rescale the radial feature following Psiformer for stability~\cite{VonGlehn_2023_PsiFormer}.
        The periodified distance follows~\cite{Cassella_2023_wigner} and is
        \begin{equation}\label{eq:periodic_distance}
            \begin{aligned}
                \norm{\vr}_{\mathbf{a}}^2 = \frac{1}{(2\pi)^2}\sum_{\mu,\nu=1}^d \Big( 
                    &\left[1 - \cos(\phi_\mu)\right] S_{\mu\nu} \left[1 - \cos(\phi_\nu)\right]\\
                    &\sin(\phi_\mu) S_{\mu\nu} \sin(\phi_\nu)\Big).
            \end{aligned}
        \end{equation}
        $\norm{\vr}_{\mathbf{a}}$ is periodic with respect to $\mathbf{a}_i$ and asymptotes to the minimum-image Euclidean distance at short range.

\begin{figure*}[t!]
    \centering
    \begin{minipage}{0.49\textwidth}
        \centering
        \includegraphics[height=0.85\textwidth]{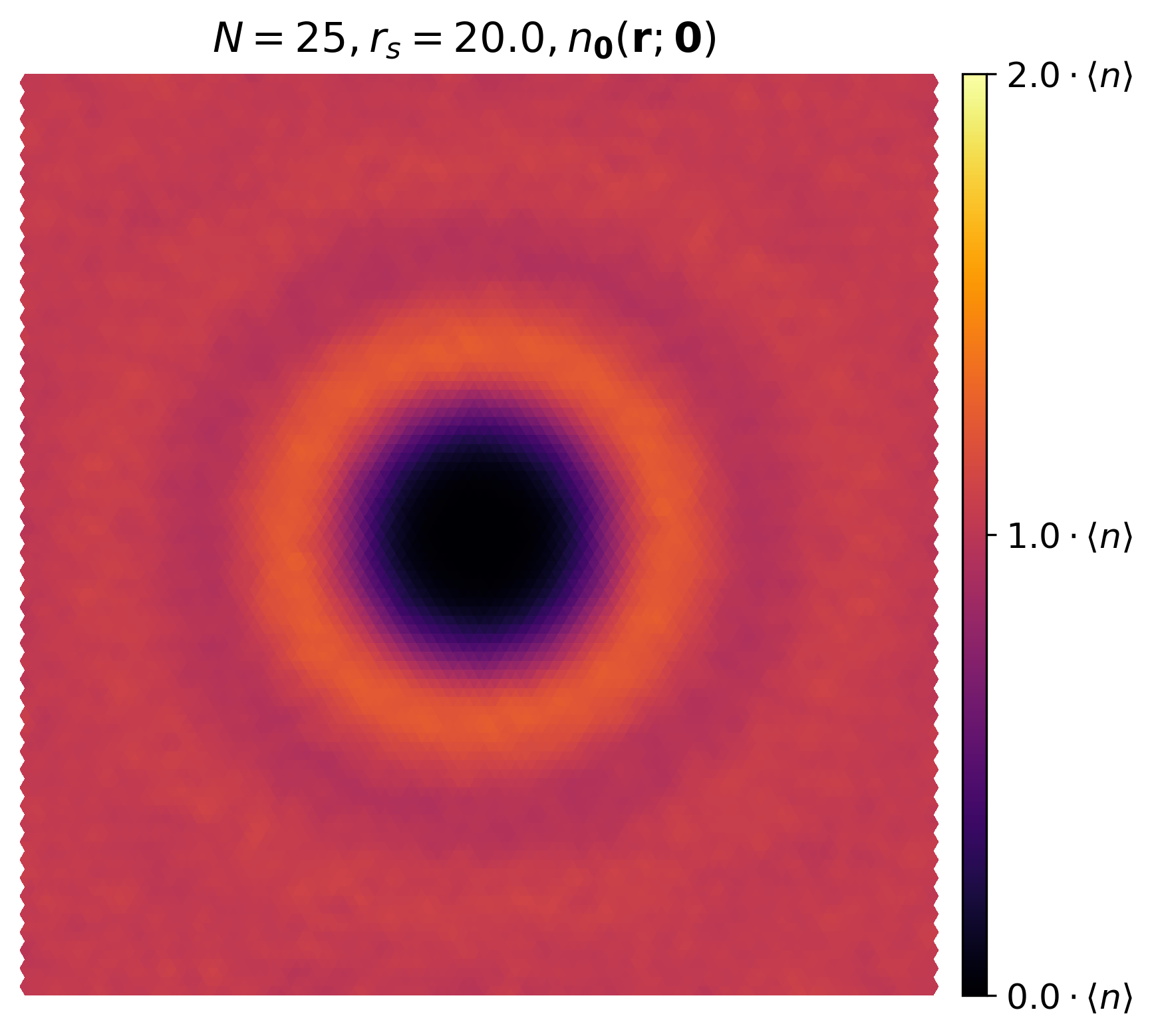}
    \end{minipage}
    \hfill
    \begin{minipage}{0.49\textwidth}
        \centering
        \includegraphics[height=0.85\textwidth]{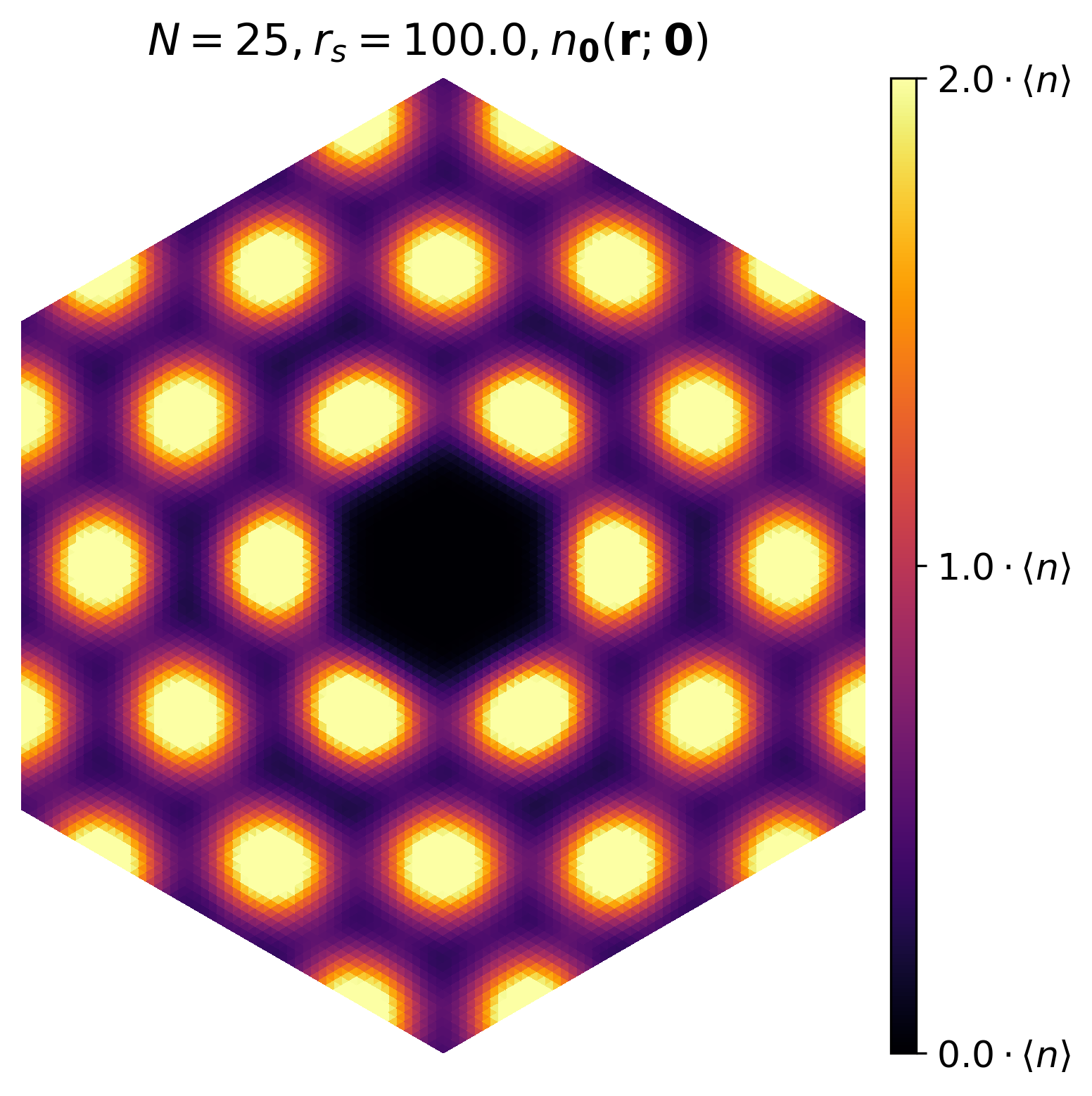}
    \end{minipage}

    \begin{minipage}{0.49\textwidth}
        \centering
        \includegraphics[height=0.85\textwidth]{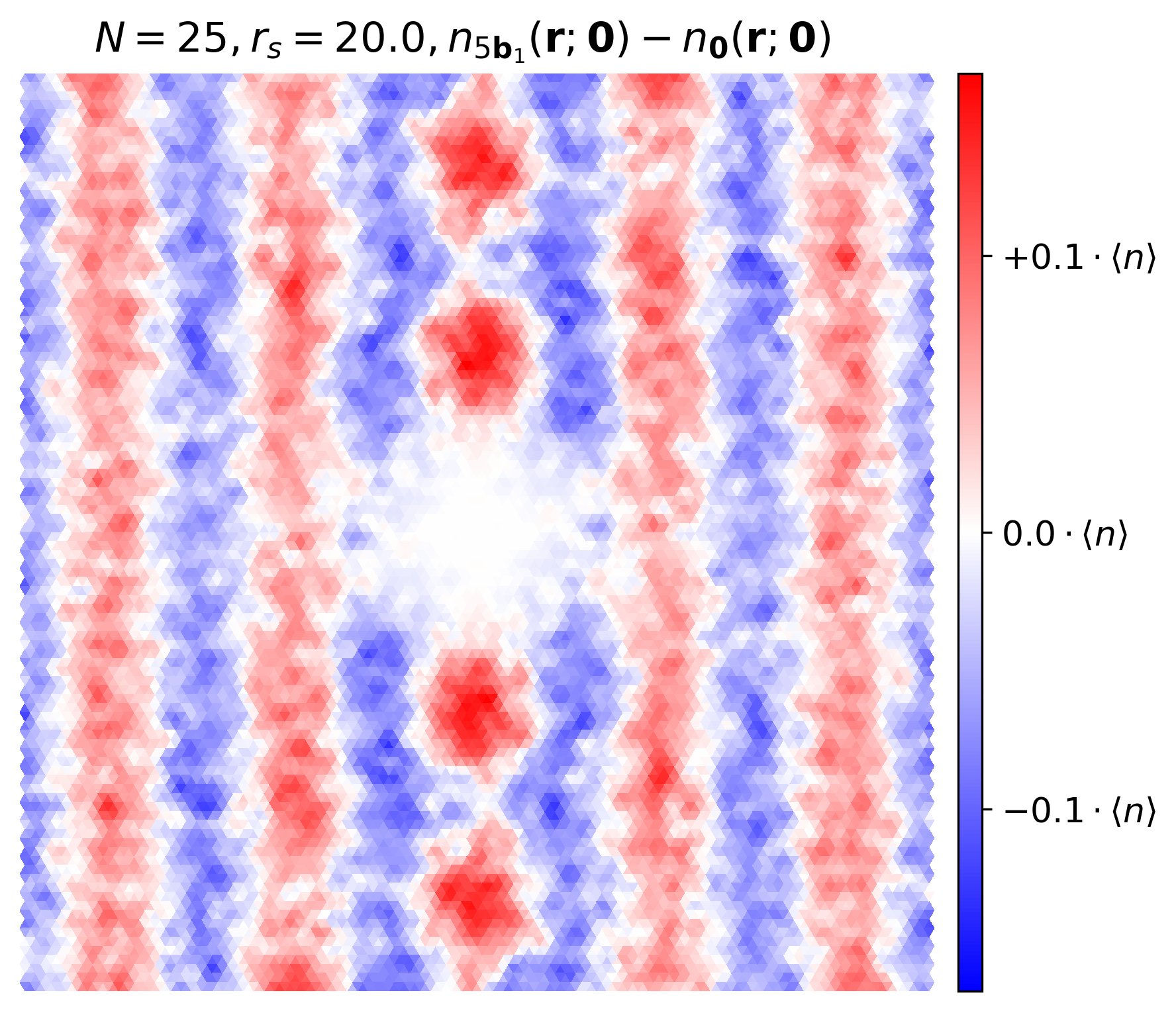}
    \end{minipage}
    \hfill
    \begin{minipage}{0.49\textwidth}
        \centering
        \includegraphics[height=0.85\textwidth]{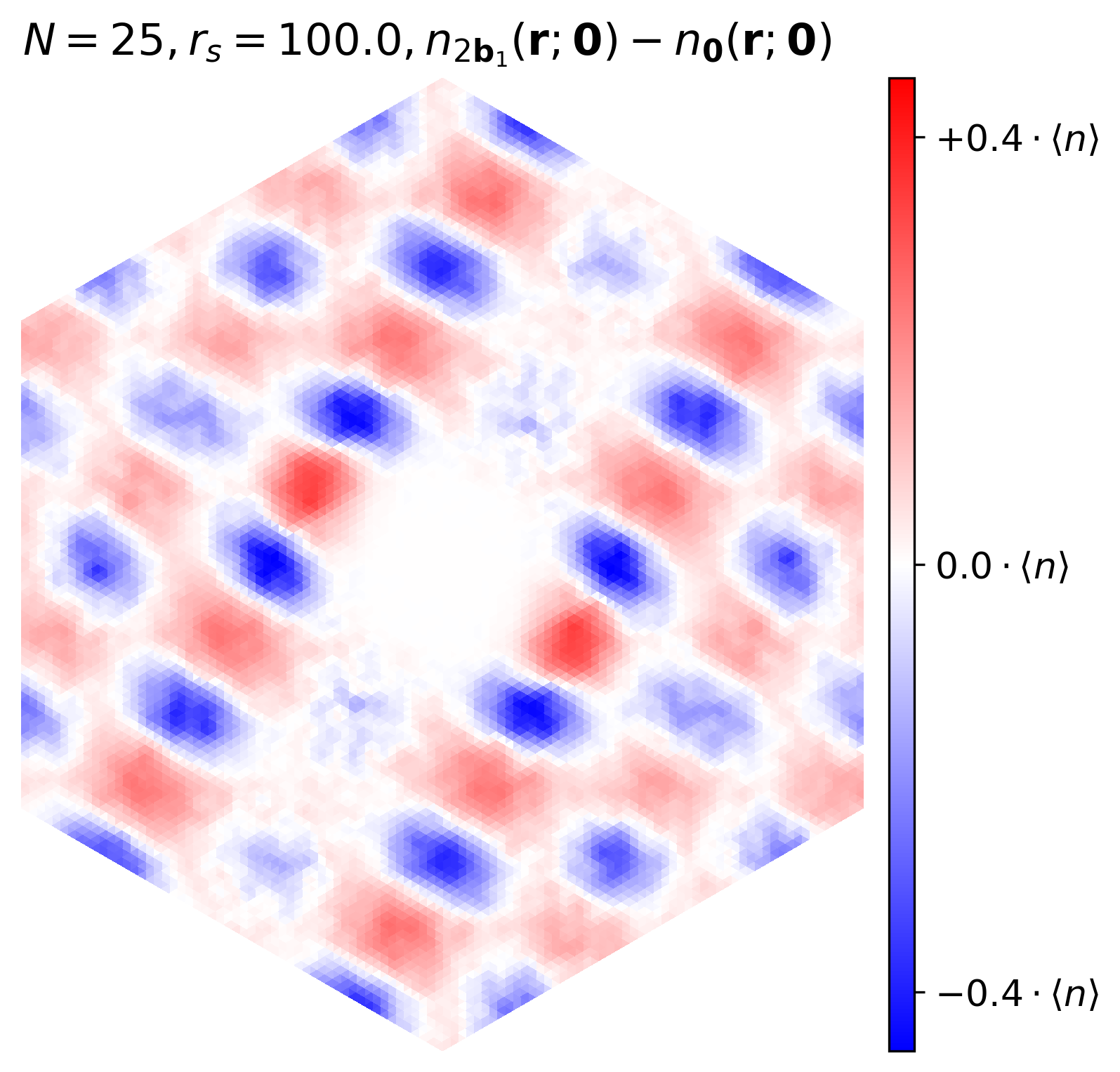}
    \end{minipage}
    
    \caption{%
    All panels show a density-density correlation function, where the conditioning particle is placed at the origin.
    Top Left: The superfluid's correlation function, which has a strong correlation hole at short distance surrounded by faint rings.
    Top Right: The Wigner crystal's correlation function, where conditional density accumulates on a sharp triangular lattice excluding the origin's correlation hole.
    Bottom Left: The roton's correlation function minus the superfluid's correlation function. The vertical accumulation and depletion bands represent a compressive excitation.
    Bottom Right: The phonon's correlation function minus the crystal's correlation function. Density depletes from the lattice sites and accumulates at the interstitials.
    }
    \label{fig:density}
\end{figure*}
        
        We construct $J$, $f$, and $g$ after $L$ EVE blocks using the final vertex and edge features, which we normalize before readout.
        $J = J_\mathrm{N} + J_\mathrm{F}$ has a learned and fixed component.
        The learned $J_\mathrm{N}$ is
        \begin{equation}
            J_\mathrm{N} \left( \{ \vr \} \right) = \mathbf{b}_\mathrm{Jastrow} \cdot \sum_{i \neq j} \vh^L_{ij}.
        \end{equation}
        The fixed $J_\mathrm{F}$ uses the same form as Psiformer~\cite{VonGlehn_2023_PsiFormer} but changes the cusp slope to be for bosons~\cite{Kato_1957_original,Drummond_2004_jastrow}.
        In Eq. \ref{eq:EVE_head}, we expect that $g$-orbitals will primarily model ground-state correlations, while $f$-orbitals will modify the correlations and imposed phase texture $e^{i \mathbf{k} \cdot \vr_i}$ for excited states.
        Therefore, because bosonic ground states of real Hamiltonians have uniform sign, we restrict $g$ to be positive but allow $f$ to be an arbitrary complex number~\footnote{The addition of $1$ is structurally a bias. We fix this because the wavefunction's overall scale is pure gauge, so the linear kernel could always be rescaled to make the bias $1$.}:
        \begin{equation}
            \begin{aligned}
                g\left(\vr_i; \{\vr_{/i}\}\right) &= \operatorname{softplus}{\left( \mathbf{b}_g \cdot \vh_i \right)},\\
                f\left(\vr_i; \{\vr_{/i}\}\right) &= 1 + \left( \mathbf{b}^\mathrm{R}_f + i\mathbf{b}^\mathrm{I}_f \right) \cdot \vh_i.\\
            \end{aligned}
        \end{equation}
        For stability, we zero-initialize $\mathbf{b}_\text{Jastrow}, \mathbf{b}_g, \mathbf{b}^\mathrm{R}_f,\mathbf{b}^\mathrm{I}_f$, so the initial wavefunction is just the fixed Jastrow and sector-enforcing phase factor.

\begin{figure*}[t!]
    \centering

    \begin{minipage}{0.49\textwidth}
        \centering
        \includegraphics[width=\linewidth]{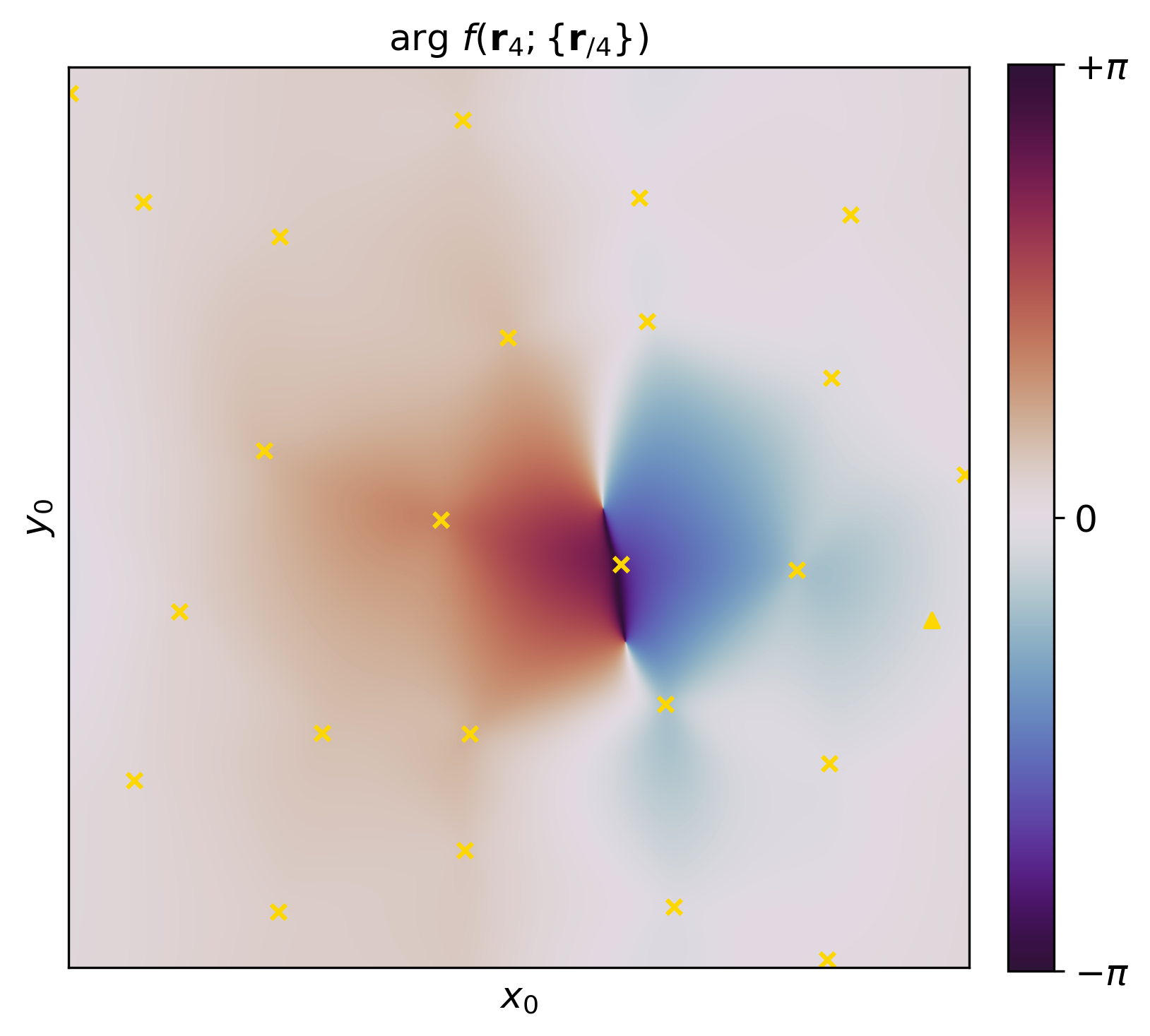}
    \end{minipage}
    \hfill
    \begin{minipage}{0.49\textwidth}
        \centering
        \includegraphics[width=\linewidth]{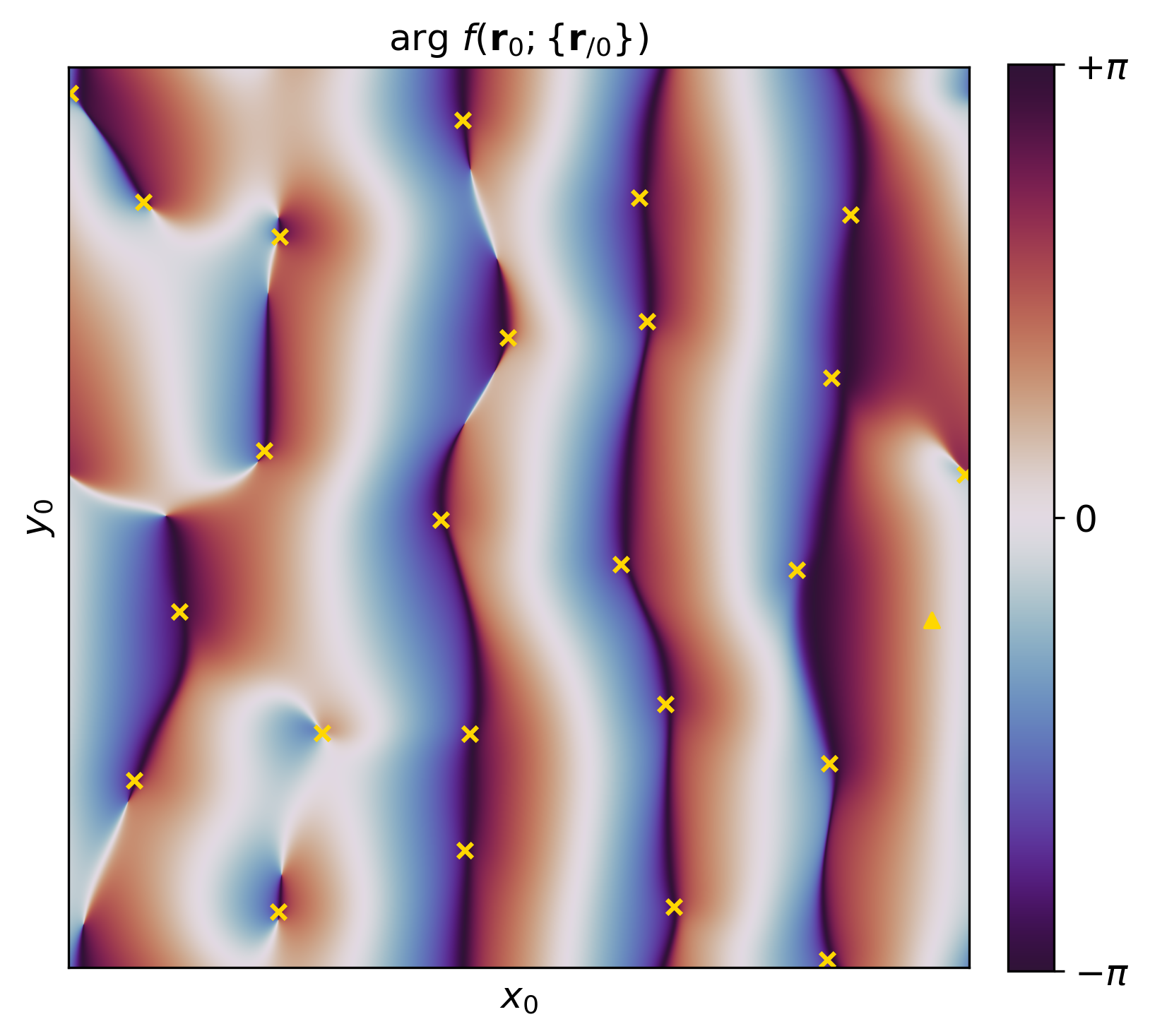}
    \end{minipage}
    \caption{%
    Left: Learned phase of $f(\vr_{4}; \{\vr_{/4}\})$ as particle $0$ is dragged around the simulation cell for the roton. The coalescence contains two counter-circulating vortices above and below.
    Right: Learned phase of $f(\vr_{0}; \{\vr_{/0}\})$ as the same particle $0$ is dragged around the simulation cell. We can see phase strings formed by merging several vertically stacked vortex pairs. Some of these even wind around the torus and are non-contractible. In both panels, crosses mark clamped particles. In both panels, note that the plotted phase excludes the fixed $e^{i\vk \cdot \vr_i}$.}
    \label{fig:phase}
\end{figure*}
    
\section{Results}

    We simulate $N=25$ bosons in the superfluid phase at $r_s = 20.0$ and in the Wigner crystal phase at $r_s = 100.0$.
    We choose these $r_s$ based on previous non-NQS quantum Monte Carlo literature~\cite{DePalo_2004_charged,Clark_2009_charged,Boninsegni_2026_plasma}.
    The superfluid simulation uses a square supercell, while the crystal simulation uses a triangular cell commensurate with a $5 \cross 5$ triangular-lattice crystal.
    In both cases, we scan momenta $\mathbf{k} = N_x \cdot \mathbf{b}_1 + 0 \cdot \mathbf{b}_2$ along the first reciprocal lattice direction.
    The appendices contain details on model configuration and training.
    
    \subsection{Momentum-Sector Energies}

    For $r_s = 20.0$, shown in Fig. \ref{fig:energy}'s top-left panel, we see a pronounced minimum near $N_x = 5$ corresponding to the roton, exactly as expected for a superfluid.
    Before the roton minimum, $E(\vk)$ rises extremely quickly, with $N_x = 2$ only slightly higher than $N_x = 1$ and $N_x = 3$ already starting to decrease.
    This could be plasmons with their $\omega \sim \sqrt{q}$ scaling mixing with two-roton states, which are not much higher in energy.
    In general, as energy increases, the interpretation of a momentum-sector ground state as a single quasiparticle becomes less clear.

    For $r_s = 100.0$, shown in Fig. \ref{fig:energy}'s top-right panel, we see pronounced zone folding indicative of crystalline order.
    To explain the dispersion, consider a pinned crystal ansatz and its translation-equivariant symmetrization
    \begin{equation}\label{eq:crystal_ansatz}
        \begin{aligned}
            \Psi^\mathrm{WC}_\mathrm{pinned}(\{\vr\}) &= \frac{1}{\sqrt{N!}}\sum_{\sigma \in S_N} \prod_{i=1}^N \phi \left( \vr_i - \vR_{\sigma_i} \right),\\
            \Psi^\mathrm{WC}_\vk(\{\vr\}) &= \int \mathrm{d}\vr' \left[ e^{i \vk \cdot \vr'} \Psi^\mathrm{WC}_\mathrm{pinned} \left( \left\{ \vr + \vr' \right\} \right) \right],\\
        \end{aligned}
    \end{equation}
    where $\phi$ is some localized function such as a Gaussian and $\vR_i$ are crystal lattice sites.
    Eq. \ref{eq:crystal_ansatz} is only nonzero if $\vk$ lies in the reciprocal lattice of the Wigner crystal, explaining the low-lying Anderson tower states at $N_x = 0,5,10,15$.
    These states have nearly the same correlations and differ mainly in their center-of-mass motion, so we expect an overall rise of $k^2/2 N r_s$.
    Around each tower state, we expect phonons, whose wavevectors extend within the emergent crystal's Brillouin zone.
    Such a quadratic-plus-sinusoid form agrees reasonably with our numerics.

    Finally, we use zero-variance extrapolation~\cite{Fu_2024_variance} to estimate the energy difference to the true momentum-sector ground states.
    Fig. \ref{fig:energy}'s bottom row shows that the relative error in the \textit{dispersion} of the highest-variance points is about $1\%$ for both superfluid and crystal, defining the dispersion's scale to be its maximum value.
    Each data point is a variational upper bound for the associated sector, and because performance is slightly worse for the higher-energy points, we would expect the quasiparticle energies to be slightly overestimated.
    Overall, this accuracy level is remarkable, given that the overall energy scale is much higher, and the physically interesting dispersion is a subtraction with large cancellations.

\begin{figure*}[t!]
        \centering
    
        \begin{minipage}{0.49\textwidth}
            \centering
            \includegraphics[width=\linewidth]{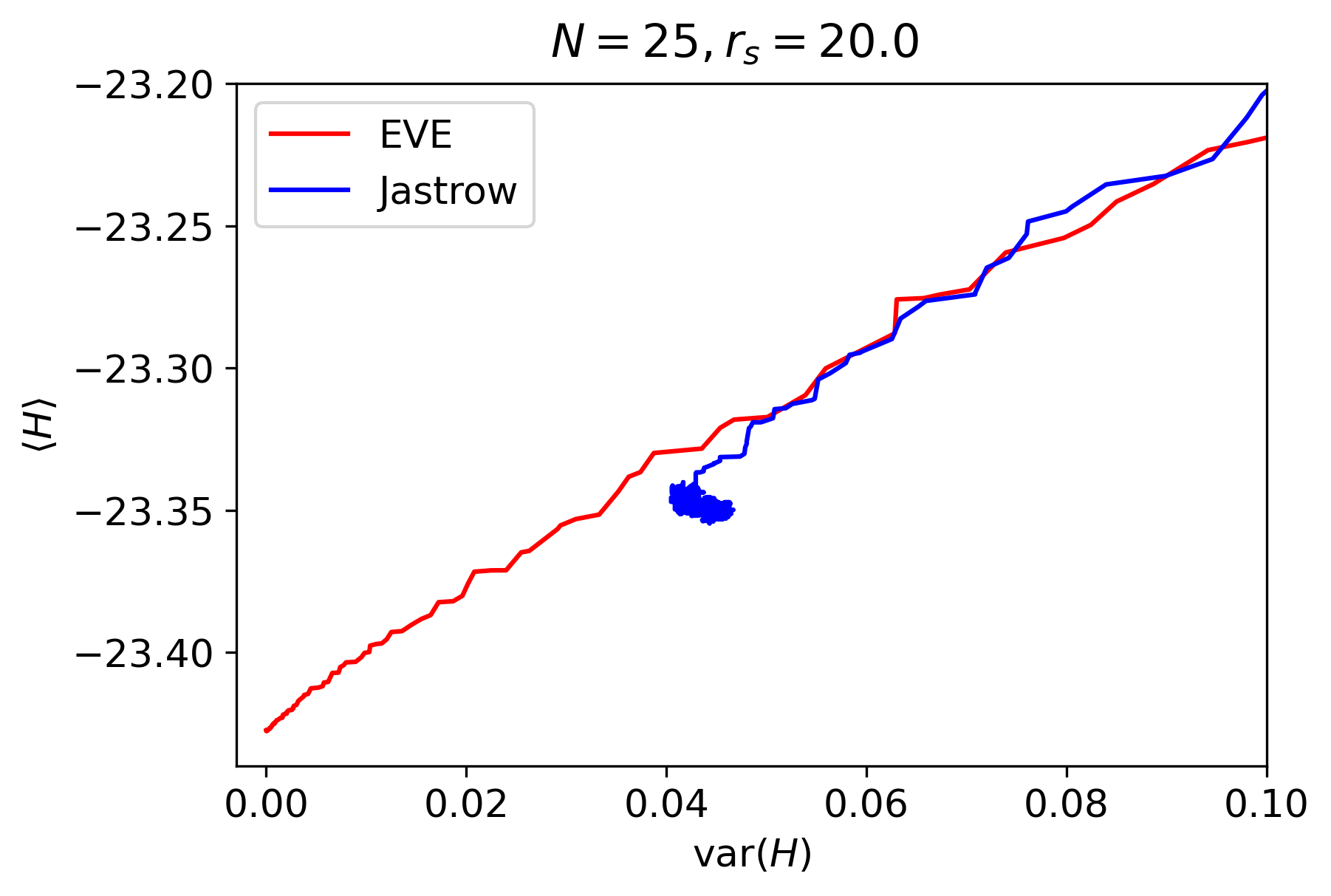}
        \end{minipage}
        \hfill
        \begin{minipage}{0.49\textwidth}
            \centering
            \includegraphics[width=\linewidth]{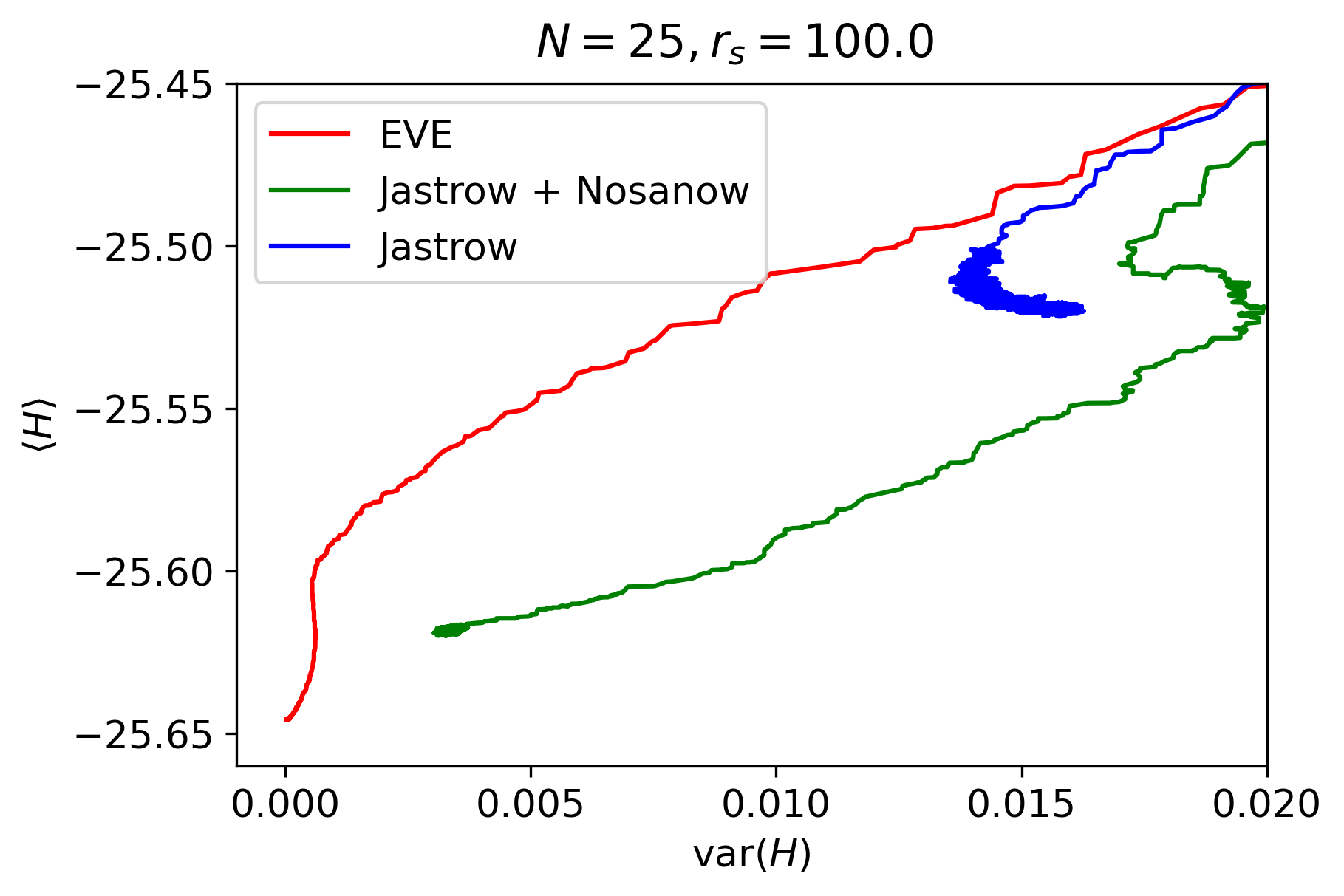}
        \end{minipage}
        \caption{Comparison of EVE to a neural-Jastrow and Nosanow-neural-Jastrow (only for the crystal) for the superfluid and Wigner crystal's global ground states. EVE has much lower energy and energy variance than the Jastrow and Nosanow-Jastrow. The plots show training measurements smoothed by taking the median over a $25$-step window.}
        \label{fig:ablations}
    \end{figure*}
        
    \subsection{Conditional Densities}

    Fig. \ref{fig:density} shows the density-density correlation $n(\vr; \mathbf{0})=\expval{n(\mathbf{r}) n(\mathbf{0})}$---the density at $\vr$ conditioned on one particle being clamped at $\mathbf{0}$---to independently diagnose the ground-state phase.
    We estimate this efficiently for our translation-invariant model by histogramming sampled $\vr_i - \vr_j$.
    Throughout, we will use density-density correlation and conditional density interchangeably.
    
    For $r_s = 20.0$, the correlation function is mostly circularly symmetric, with a strong short-range hole surrounded by faint rings, exactly as expected for a liquid.
    We also verified superfluidity by checking the condensate fraction, which is about $16\%$.
    For $r_s = 100.0$, crystalline order is apparent and striking---beyond the central hole, density accumulates cleanly on a triangular lattice.
    The density between lattice peaks is suppressed but not vanishing, demonstrating that quantum effects are still active.
    We emphasize that the plot is of a correlation function; the unconditional density would be constant.

    To probe the excited states' correlation structure, we also plot their $n(\vr; \mathbf{0})$ with the global ground-state $n(\vr; \mathbf{0})$ subtracted off for contrast.
    For the roton, there are clear bands of density accumulation and depletion oriented perpendicular to the wavevector, indicating a longitudinal compressive excitation.
    For the $\vk = 2\mathbf{b}_1$ phonon, there is again a banded pattern perpendicular to the phonon wavevector, but also significant variation along the perpendicular direction.
    In general, density depletes from the triangular lattice sites and accumulates at the interstitials.
    The pattern is consistent with a transverse excitation and certainly looks different from the roton, but we cannot make a definitive diagnosis.

    \subsection{Roton Visualization}

    We visualize the roton's phase structure and assess how well the ``microscopic vortex ring'' intuition~\cite{Feynman_1956_backflow} describes it.
    In an $N$-particle system for $N \gg 1$, on average most particles are not participating in the quasiparticle, so examining the many-body wavefunction itself is not fruitful.
    Instead, we visualize the orbital $f\left(\vr_i; \{\vr_{/i}\}\right)$, which is the learned phase imparted to particle $i$ when it is the particle boosted by the exponential phase factor.

    To form a 2D visualization, we sample a configuration of $N$ particles, clamp $N-1$ of them, and drag the remaining particle around the simulation cell.
    In Fig. \ref{fig:phase}'s left panel, the dragged particle is not the same as the boosted particle.
    Here, the learned phase is only significant when the dragged particle nears the boosted particle, at which point it experiences a pair of opposite-circulation vortices, the 2D analog of the 3D vortex ring.
    
    In Fig. \ref{fig:phase}'s right panel, the dragged and boosted particles are the same, so the learned phase is significant near every coalescence, and the vortex polarities reverse.
    The separation between the two vortices is comparable to the interparticle distance, and we can see groups of vertically stacked particles whose dipoles merge into continuous phase strings.
    This partial cancellation of the vortices, which only leaves a vortex at each end of the string connected by a phase slip (or just a phase slip if the string wraps around the torus), may lower energy.

    Thus, while the phase dipole intuition has an element of truth, it does not account for the finite spatial separation between the phase vortices or the higher-order structures formed by cooperating groups of particles.
    This demonstrates the value of a fully generic momentum-$\vk$ ansatz not bound by excessive backflow priors.

    \subsection{Comparison to Jastrow Wavefunction}

    We compare results to a Jastrow and Nosanow-Jastrow wavefunction as a sanity check.
    We parameterize the Jastrow as a neural network to approach the best energy over all pair-product-type wavefunctions, loosely inspired by~\cite{Geier_2025_moire}.
    We use the same input features, model dimension, and number of blocks as EVE.
    Each block is $\vh_{ij} \leftarrow \vh_{ij} + \MLP{(\RMSNorm{(\vh_{ij})})}$.
    After the final block, we use the same learned-plus-fixed readout as the main EVE model.
    For the solid, we also test including a Nosanow factor, which is a distinguishable product over orbitals~\cite{Nosanow_1962_hartree,Magro_1994_log}.
    The Nosanow breaks permutation symmetry but cannot cheat the energy lower, because the true ground state is permutation-symmetric anyway.

    For both $r_s = 20.0$ and $r_s = 100.0$, the Jastrow's energy and energy variance are significantly higher than EVE's as shown in Fig. \ref{fig:ablations}.
    Furthermore, at $r_s = 100.0$, probing the conditional density finds that the Jastrow itself fails to find crystalline order.
    Including the Nosanow factor improves results significantly over a pure Jastrow but is still very far from EVE.

\section{Discussion}

    In this work, we proposed EVE, the first NQS architecture for continuum particles that is also a momentum eigenstate with configurable momentum.
    Using EVE, we probed the momentum-sector ground states of 2D bosons with mutual $1/r$ repulsion, finding that a single unified ansatz with minimal human biases can capture four qualitatively different wavefunctions: liquid, roton, crystal, and phonon.
    In both the superfluid regime at $r_s = 20.0$ and Wigner crystal regime at $r_s = 100.0$, we diagnosed the underlying phase of matter through $E(\vk)$'s striking shape---a single roton minimum for the superfluid, and a periodic-plus-tower structure for the crystal.
    By measuring density-density correlation functions, we both confirmed the ground-state phase diagnoses and probed the excitations' correlation structure.
    
    In future work, we plan to explore extending EVE to fermionic systems.
    Here, we choose bosons primarily because of the famous roton minimum, which provides a natural test of the architecture.
    For fermions, we anticipate that the main change would be the output head; prior to orbital projection, the equivariant backbone can probably remain the same.
    Given our success on the floating crystal, we also plan to test EVE's competitiveness against non-invariant architectures on ground-state problems.
    We see two potential advantages.
    First, because EVE treats two-particle features as ``first-class'', it may be capable of learning correlations faster than architectures which use only one-particle inputs.
    Second, the model does not need to learn the ground state's symmetries because they are directly built in, which may further accelerate convergence, especially for observables other than the energy.

    By building in the target momentum sector in the architecture rather than in training, EVE makes quasiparticles as easy as the ground state.
    Conceptually, EVE introduces a new paradigm in NQS that reconciles architectural flexibility with known physical constraints.
    On one hand, EVE's flexible and unbiased design reflects modern deep learning practice~\cite{Sutton_2019_bitter}, which has observed empirically that as compute grows, flexible and scalable models outperform models encumbered by excessive human priors.
    On the other hand, by strictly enforcing known and provable properties of energy eigenstates, EVE completes qualitatively new tasks, such as targeting momentum-sector ground states.
    We predict that EVE will open many doors in NQS, both among the physical problems that NQS can solve and in the design of NQS themselves.

\section*{Acknowledgments}
    DDD was supported by the National Science Foundation Graduate Research Fellowship under Grant No. 2141064.
    The authors acknowledge the MIT SuperCloud, Lincoln Laboratory Supercomputing Center (LLSC), and MIT Office of Research Computing and Data (ORCD) for providing high-performance computing resources.
    This material is based upon work supported in part by the MIT Institute for Soldier Nanotechnologies. 
    This work is supported by the National Science Foundation under Cooperative Agreement PHY-2019786 (The NSF AI Institute for Artificial Intelligence and Fundamental Interactions, http://iaifi.org/).

\appendix

\section{Training Details}

    For sampling, we implemented a rotating refresh carousel to improve the Markov chain's global mixing and decorrelation.
    In addition to the primary walker batch that we use for gradient updates, we maintain a parallel auxiliary batch.
    After every training step, we randomly select and discard a small fraction of the primary batch and replace it with walkers retrieved from the auxiliary batch.
    In the auxiliary batch, retrieved walkers are replaced by re-initialized walkers sampled from the uniform distribution.
    The retrieved portion of the auxiliary batch rotates, such that between an auxiliary walker being initialized and retrieved, exactly $t_\mathrm{burn}$ gradient updates pass.
    Furthermore, each primary batch walker lives for on average $t_\mathrm{refresh}$ gradient updates before refresh.
    The refresh carousel prevents walkers from trapping at local minima in the probability distribution and helps defend against spurious modes~\cite{Goodfellow_2016_book,Westerhout_2020_generalize,Zhang_2023_spurious}.

    We train using stochastic reconfiguration (SR)~\cite{Sorella_1998_green,Sorella_2001_lanczos}, also known as natural gradient descent~\cite{Amari_1998_natural}, using the efficient sample-space formulation~\cite{Chen_2024_miNSR,Rende_2024_simple}.
    We use Ewald summation and a neutralizing background to handle the $1/r$ interaction's long-range tail, and we use forward Laplacian to accelerate local energy calculation~\cite{Li_2024_folx}.
    We use the same local energy clipping as Psiformer~\cite{VonGlehn_2023_PsiFormer}.
    After SR-preconditioning, we only use simple stochastic gradient descent without momentum or norm constraint.

\section{Detailed Configuration}

    We use vertex dimension $d_1 = 128$, edge dimension $d_2 = 32$, $n_\mathrm{heads} = 4$ edge-attention heads, MLP width factors of $1.0$ (the MLP's hidden dimension is the same as the input feature dimension), and $L = 4$ EVE blocks.
    The total parameter count is a modest $\num{198850}$ and is independent of particle number.
    For the ground state, the $f$-orbital is disabled, because bosonic ground states have uniform sign everywhere.
    Following the observations in~\cite{Dai_2026_barrier}, we freeze the Jastrow's $\alpha$ (see the original Psiformer~\cite{VonGlehn_2023_PsiFormer}) and feature $\alpha_\mathrm{soft}$ to $0.1$ for the superfluid at $r_s = 20.0$ and $0.2$ for the Wigner crystal at $r_s = 100.0$.
    Results are not particularly sensitive to $\alpha$'s exact value, so clamping them reduces the number of parameters with disproportionate influence on the wavefunction.

    Sampling uses a batch size of $\num{2048}$, $50$ propose-accept/reject steps between gradient updates, $t_\mathrm{refresh} = 100$, and $t_\mathrm{burn} = 100$.
    The proposal distribution is an isotropic Gaussian over all particles, where the proposal width is adjusted every $50$ proposals by multiplying or dividing by $1.01$ to keep the acceptance probability near $0.234$.
    After completing training, we infer results using freshly burned-in walkers that the model has never seen before.
    Inference burn in uses $500 \times 50$ propose-accept/reject steps.
    To eliminate autocorrelation, we only average across the inference batch of $2^{14} = \num{16384}$ walkers and not over time.
    We do not clip or discard outliers during inference.
    We estimate uncertainties using jackknife resampling~\cite{Efron_1994_jackknife}.

    For all simulations, we use an SR diagonal shift of $10^{-4}$ and clipping factor of $64.0$.
    Our learning rate schedule consists of a linear warmup over $\num{500}$ steps to the peak learning rate followed by a linear anneal over $\num{9500}$ steps down to zero.
    Results can be modestly improved by training for longer, but $\num{10000}$ steps reduced cost and already achieved acceptable results.
    A peak learning rate of $0.1$ worked for all but the largest momenta, for which we had to reduce learning rate, likely due to the increasing kinetic energy scale.
    
    Except for the aforementioned high momenta, all sectors were trained independently; the highest-momenta sectors used pretraining, i.e. starting the training using the final parameters of a completed lower-momentum run.
    For the liquid, we started pretraining at $N_x = 9$, i.e. the $N_x = 9$ run was warm-started with $N_x = 8$'s final checkpoint.
    $N_x = 0.9$ used peak learning rate $0.03$, and $N_x = 10.0$ used peak learning rate $0.01$.
    For the solid, we started pretraining at $N_x = 13$.
    $N_x = 13$ used peak learning rate $0.05$, $N_x = 14$ used peak learning rate $0.03$, and $N_x = 15$ used peak learning rate $0.03$.

\section{Comparison to Other Architectures}

    EVE's overall setup is inspired by FermiNet~\cite{Pfau_2020_FermiNet}, which also uses parallel one-body (vertex) and two-body (edge) streams and a two-body to one-body aggregation update. However, FermiNet's aggregation is a nonselective mean-pool, while we use learned edge attention, and unlike FermiNet we also allow vertices to update edges.
    The idea of using pairwise features to control aggregation was inspired by SchNet~\cite{Schutt_2017_SchNet,Gerard_2022_gold}, but they are convolution-like instead of attention-like.

    Other architectures that operate on fully connected graphs include the message-passing NQS (MP-NQS)~\cite{Pescia_2024_MPNQS} and GemiNet~\cite{Luo_2026_GemiNet}.
    MP-NQS has an $\order{N^3}$ forward pass due to a contraction-over-particles in their particle attention mechanism (see Eq. 8 in ~\cite{Pescia_2024_MPNQS}), while ours is only $\order{N^2}$, and MP-NQS is a backflow architecture using hand-chosen orbitals.
    MP-NQS has about one-tenth the parameter count of EVE due to its more specific priors, so it is a much more compact representation, but parameter count and runtime cost are not necessarily related, and modern optimization techniques such as minSR~\cite{Chen_2024_miNSR} mean that large parameter counts are no longer the bottlenecks that they once were.
    The MP-NQS authors mention the possibility of modeling the homogeneous electron gas at fixed momentum~\cite{Pescia_2024_MPNQS}, but this requires backflow around plane-wave orbitals specifically, and their reported results for the Wigner crystal use site-centered Gaussians.
    In any case, even if the possibility exists, the MP-NQS paper only reports ground state results.
    GemiNet is $\order{N^2}$ like us, but both the actual update designs and physical settings are very different.

\bibliography{references}

@article{Landau_1941_superfluidity,
    title = {Theory of the Superfluidity of Helium {II}},
    author = {Landau, L.},
    journal = {Phys. Rev.},
    volume = {60},
    issue = {4},
    pages = {356--358},
    numpages = {0},
    year = {1941},
    month = {Aug},
    publisher = {American Physical Society},
    doi = {10.1103/PhysRev.60.356},
    url = {https://link.aps.org/doi/10.1103/PhysRev.60.356}
}

@article{Feynman_1954_theory,
    title = {Atomic Theory of the Two-Fluid Model of Liquid Helium},
    author = {Feynman, R. P.},
    journal = {Phys. Rev.},
    volume = {94},
    issue = {2},
    pages = {262--277},
    numpages = {0},
    year = {1954},
    month = {Apr},
    publisher = {American Physical Society},
    doi = {10.1103/PhysRev.94.262},
    url = {https://link.aps.org/doi/10.1103/PhysRev.94.262}
}

@article{Feynman_1956_backflow,
    title = {Energy Spectrum of the Excitations in Liquid Helium},
    author = {Feynman, R. P. and Cohen, Michael},
    journal = {Phys. Rev.},
    volume = {102},
    issue = {5},
    pages = {1189--1204},
    numpages = {0},
    year = {1956},
    month = {Jun},
    publisher = {American Physical Society},
    doi = {10.1103/PhysRev.102.1189},
    url = {https://link.aps.org/doi/10.1103/PhysRev.102.1189}
}

@article{Nambu_1960_quasiparticle,
    title = {Quasi-Particles and Gauge Invariance in the Theory of Superconductivity},
    author = {Nambu, Yoichiro},
    journal = {Phys. Rev.},
    volume = {117},
    issue = {3},
    pages = {648--663},
    numpages = {0},
    year = {1960},
    month = {Feb},
    publisher = {American Physical Society},
    doi = {10.1103/PhysRev.117.648},
    url = {https://link.aps.org/doi/10.1103/PhysRev.117.648}
}

@article{Goldstone_1961_field_theory,
    title={Field theories with «superconductor» solutions},
    volume={19},
    DOI={10.1007/bf02812722},
    number={1},
    journal={Il Nuovo Cimento},
    author={Goldstone, J.},
    year={1961},
    month={Jan},
    pages={154–164},
}

@article{Goldstone_1962_broken,
    title = {Broken Symmetries},
    author = {Goldstone, Jeffrey and Salam, Abdus and Weinberg, Steven},
    journal = {Phys. Rev.},
    volume = {127},
    issue = {3},
    pages = {965--970},
    numpages = {0},
    year = {1962},
    month = {Aug},
    publisher = {American Physical Society},
    doi = {10.1103/PhysRev.127.965},
    url = {https://link.aps.org/doi/10.1103/PhysRev.127.965}
}

@article{Tsui_1982_experiment,
    title = {Two-Dimensional Magnetotransport in the Extreme Quantum Limit},
    author = {Tsui, D. C. and Stormer, H. L. and Gossard, A. C.},
    journal = {Phys. Rev. Lett.},
    volume = {48},
    issue = {22},
    pages = {1559--1562},
    numpages = {0},
    year = {1982},
    month = {May},
    publisher = {American Physical Society},
    doi = {10.1103/PhysRevLett.48.1559},
    url = {https://link.aps.org/doi/10.1103/PhysRevLett.48.1559}
}

@article{Laughlin_1983_theory,
    title = {Anomalous Quantum Hall Effect: An Incompressible Quantum Fluid with Fractionally Charged Excitations},
    author = {Laughlin, R. B.},
    journal = {Phys. Rev. Lett.},
    volume = {50},
    issue = {18},
    pages = {1395--1398},
    numpages = {0},
    year = {1983},
    month = {May},
    publisher = {American Physical Society},
    doi = {10.1103/PhysRevLett.50.1395},
    url = {https://link.aps.org/doi/10.1103/PhysRevLett.50.1395}
}

@article{Arovas_1984_statistics,
    title = {Fractional Statistics and the Quantum Hall Effect},
    author = {Arovas, Daniel and Schrieffer, J. R. and Wilczek, Frank},
    journal = {Phys. Rev. Lett.},
    volume = {53},
    issue = {7},
    pages = {722--723},
    numpages = {0},
    year = {1984},
    month = {Aug},
    publisher = {American Physical Society},
    doi = {10.1103/PhysRevLett.53.722},
    url = {https://link.aps.org/doi/10.1103/PhysRevLett.53.722}
}

@article{Moore_1991_nonabelian,
    title = {Nonabelions in the fractional quantum hall effect},
    journal = {Nuclear Physics B},
    volume = {360},
    number = {2},
    pages = {362-396},
    year = {1991},
    issn = {0550-3213},
    doi = {https://doi.org/10.1016/0550-3213(91)90407-O},
    url = {https://www.sciencedirect.com/science/article/pii/055032139190407O},
    author = {Gregory Moore and Nicholas Read},
}

@article{Wen_1990_order,
    author = {Wen, X. G.},
    title = {TOPOLOGICAL ORDERS IN RIGID STATES},
    journal = {International Journal of Modern Physics B},
    volume = {04},
    number = {02},
    pages = {239-271},
    year = {1990},
    doi = {10.1142/S0217979290000139},
    URL = {https://doi.org/10.1142/S0217979290000139},
}

@article{Wen_1990_degeneracy,
    title = {Ground-state degeneracy of the fractional quantum Hall states in the presence of a random potential and on high-genus Riemann surfaces},
    author = {Wen, X. G. and Niu, Q.},
    journal = {Phys. Rev. B},
    volume = {41},
    issue = {13},
    pages = {9377--9396},
    numpages = {0},
    year = {1990},
    month = {May},
    publisher = {American Physical Society},
    doi = {10.1103/PhysRevB.41.9377},
    url = {https://link.aps.org/doi/10.1103/PhysRevB.41.9377}
}

@article{Zhao_2016_shift,
    title={An efficient variational principle for the direct optimization of excited states},
    volume={12},
    DOI={10.1021/acs.jctc.6b00508},
    number={8},
    journal={Journal of Chemical Theory and Computation},
    author={Zhao, Luning and Neuscamman, Eric},
    year={2016},
    month={Jul},
    pages={3436–3440},
}

@article{Choo_2018_orthogonalize,
    title = {Symmetries and Many-Body Excitations with Neural-Network Quantum States},
    author = {Choo, Kenny and Carleo, Giuseppe and Regnault, Nicolas and Neupert, Titus},
    journal = {Phys. Rev. Lett.},
    volume = {121},
    issue = {16},
    pages = {167204},
    numpages = {6},
    year = {2018},
    month = {Oct},
    publisher = {American Physical Society},
    doi = {10.1103/PhysRevLett.121.167204},
    url = {https://link.aps.org/doi/10.1103/PhysRevLett.121.167204}
}

@article{Pathak_2021_penalty,
    author = {Pathak, Shivesh and Busemeyer, Brian and Rodrigues, João N. B. and Wagner, Lucas K.},
    title = {Excited states in variational {Monte Carlo} using a penalty method},
    journal = {The Journal of Chemical Physics},
    volume = {154},
    number = {3},
    pages = {034101},
    year = {2021},
    month = {01},
    issn = {0021-9606},
    doi = {10.1063/5.0030949},
    url = {https://doi.org/10.1063/5.0030949},
}

@article{Entwistle_2023_penalty,
    title={Electronic excited states in deep variational {Monte Carlo}},
    volume={14},
    DOI={10.1038/s41467-022-35534-5},
    number={1},
    eid = {274},
    journal={Nature Communications},
    author={Entwistle, M. T. and Schätzle, Z. and Erdman, P. A. and Hermann, J. and Noé, F.},
    year={2023},
    month={Jan},
}

@article{Wheeler_2024_ensemble,
    doi = {10.1088/2516-1075/ad38f8},
    url = {https://doi.org/10.1088/2516-1075/ad38f8},
    year = {2024},
    month = {apr},
    publisher = {IOP Publishing},
    volume = {6},
    number = {2},
    pages = {025001},
    author = {Wheeler, William A and Kleiner, Kevin G and Wagner, Lucas K},
    title = {Ensemble variational {Monte Carlo} for optimization of correlated excited state wave functions},
    journal = {Electronic Structure},
}

@article{Pfau_2024_excited,
    author = {David Pfau  and Simon Axelrod  and Halvard Sutterud  and Ingrid von Glehn  and James S. Spencer },
    title = {Accurate computation of quantum excited states with neural networks},
    journal = {Science},
    volume = {385},
    number = {6711},
    pages = {eadn0137},
    year = {2024},
    doi = {10.1126/science.adn0137},
    URL = {https://www.science.org/doi/abs/10.1126/science.adn0137},
}

@article{
    Romero_2025_spectroscopy,
    title={Spectroscopy of two-dimensional interacting lattice electrons using symmetry-aware neural backflow transformations},
    volume={8},
    DOI={10.1038/s42005-025-01955-z},
    number={1},
    journal={Communications Physics},
    author={Romero, Imelda and Nys, Jannes and Carleo, Giuseppe},
    year={2025},
    month={Jan},
}

@article{McMillan_1965_ground,
    title = {Ground State of Liquid {${\mathrm{He}}^{4}$}},
    author = {McMillan, W. L.},
    journal = {Phys. Rev.},
    volume = {138},
    issue = {2A},
    pages = {A442--A451},
    numpages = {0},
    year = {1965},
    month = {Apr},
    publisher = {American Physical Society},
    doi = {10.1103/PhysRev.138.A442},
    url = {https://link.aps.org/doi/10.1103/PhysRev.138.A442}
}

@article{Ceperley_1977_first,
    title = {{M}onte {C}arlo simulation of a many-fermion study},
    author = {Ceperley, D. and Chester, G. V. and Kalos, M. H.},
    journal = {Phys. Rev. B},
    volume = {16},
    issue = {7},
    pages = {3081--3099},
    numpages = {0},
    year = {1977},
    month = {Oct},
    publisher = {American Physical Society},
    doi = {10.1103/PhysRevB.16.3081},
    url = {https://link.aps.org/doi/10.1103/PhysRevB.16.3081}
}

@article{Ceperley_1980_electron,
    title = {Ground State of the Electron Gas by a Stochastic Method},
    author = {Ceperley, D. M. and Alder, B. J.},
    journal = {Phys. Rev. Lett.},
    volume = {45},
    issue = {7},
    pages = {566--569},
    numpages = {0},
    year = {1980},
    month = {Aug},
    publisher = {American Physical Society},
    doi = {10.1103/PhysRevLett.45.566},
    url = {https://link.aps.org/doi/10.1103/PhysRevLett.45.566}
}

@article{Tanatar_1989_2DEG,
    title = {Ground state of the two-dimensional electron gas},
    author = {Tanatar, B. and Ceperley, D. M.},
    journal = {Phys. Rev. B},
    volume = {39},
    issue = {8},
    pages = {5005--5016},
    numpages = {0},
    year = {1989},
    month = {Mar},
    publisher = {American Physical Society},
    doi = {10.1103/PhysRevB.39.5005},
    url = {https://link.aps.org/doi/10.1103/PhysRevB.39.5005}
}

@article{Foulkes_2001_review,
    title = {Quantum {M}onte {C}arlo simulations of solids},
    author = {Foulkes, W. M. C. and Mitas, L. and Needs, R. J. and Rajagopal, G.},
    journal = {Rev. Mod. Phys.},
    volume = {73},
    issue = {1},
    pages = {33--83},
    numpages = {0},
    year = {2001},
    month = {Jan},
    publisher = {American Physical Society},
    doi = {10.1103/RevModPhys.73.33},
    url = {https://link.aps.org/doi/10.1103/RevModPhys.73.33}
}

@article{Needs_2010_review,
    doi = {10.1088/0953-8984/22/2/023201},
    url = {https://doi.org/10.1088/0953-8984/22/2/023201},
    year = {2009},
    month = {dec},
    publisher = {},
    volume = {22},
    number = {2},
    pages = {023201},
    author = {Needs, R J and Towler, M D and Drummond, N D and López Ríos, P},
    title = {Continuum variational and diffusion quantum {M}onte {C}arlo calculations},
    journal = {Journal of Physics: Condensed Matter},
}

@article{Carleo_2017_solving,
    author = {Giuseppe Carleo  and Matthias Troyer },
    title = {Solving the quantum many-body problem with artificial neural networks},
    journal = {Science},
    volume = {355},
    number = {6325},
    pages = {602-606},
    year = {2017},
    doi = {10.1126/science.aag2302},
    URL = {https://www.science.org/doi/abs/10.1126/science.aag2302},
}

@article{Nomura_2017_RBM,
    title = {Restricted {B}oltzmann machine learning for solving strongly correlated quantum systems},
    author = {Nomura, Yusuke and Darmawan, Andrew S. and Yamaji, Youhei and Imada, Masatoshi},
    journal = {Phys. Rev. B},
    volume = {96},
    issue = {20},
    pages = {205152},
    numpages = {8},
    year = {2017},
    month = {Nov},
    publisher = {American Physical Society},
    doi = {10.1103/PhysRevB.96.205152},
    url = {https://link.aps.org/doi/10.1103/PhysRevB.96.205152}
}

@article{Carleo_2019_NetKet,
    title = {{N}et{K}et: {A} machine learning toolkit for many-body quantum systems},
    journal = {SoftwareX},
    volume = {10},
    pages = {100311},
    year = {2019},
    issn = {2352-7110},
    doi = {https://doi.org/10.1016/j.softx.2019.100311},
    url = {https://www.sciencedirect.com/science/article/pii/S2352711019300974},
    author = {Giuseppe Carleo and Kenny Choo and Damian Hofmann and James E.T. Smith and Tom Westerhout and Fabien Alet and Emily J. Davis and Stavros Efthymiou and Ivan Glasser and Sheng-Hsuan Lin and Marta Mauri and Guglielmo Mazzola and Christian B. Mendl and Evert {van Nieuwenburg} and Ossian O’Reilly and Hugo Théveniaut and Giacomo Torlai and Filippo Vicentini and Alexander Wietek},

}

@article{Choo_2019_CNN,
    title = {Two-dimensional frustrated ${J}_{1}\text{\ensuremath{-}}{J}_{2}$ model studied with neural network quantum states},
    author = {Choo, Kenny and Neupert, Titus and Carleo, Giuseppe},
    journal = {Phys. Rev. B},
    volume = {100},
    issue = {12},
    pages = {125124},
    numpages = {7},
    year = {2019},
    month = {Sep},
    publisher = {American Physical Society},
    doi = {10.1103/PhysRevB.100.125124},
    url = {https://link.aps.org/doi/10.1103/PhysRevB.100.125124}
}

@article{Sharir_2020_RNN,
    title = {Deep Autoregressive Models for the Efficient Variational Simulation of Many-Body Quantum Systems},
    author = {Sharir, Or and Levine, Yoav and Wies, Noam and Carleo, Giuseppe and Shashua, Amnon},
    journal = {Phys. Rev. Lett.},
    volume = {124},
    issue = {2},
    pages = {020503},
    numpages = {6},
    year = {2020},
    month = {Jan},
    publisher = {American Physical Society},
    doi = {10.1103/PhysRevLett.124.020503},
    url = {https://link.aps.org/doi/10.1103/PhysRevLett.124.020503}
}

@article{Pfau_2020_FermiNet,
    title = {Ab initio solution of the many-electron {S}chr\"odinger equation with deep neural networks},
    author = {Pfau, David and Spencer, James S. and Matthews, Alexander G. D. G. and Foulkes, W. M. C.},
    journal = {Phys. Rev. Res.},
    volume = {2},
    issue = {3},
    pages = {033429},
    numpages = {20},
    year = {2020},
    month = {Sep},
    publisher = {American Physical Society},
    doi = {10.1103/PhysRevResearch.2.033429},
    url = {https://link.aps.org/doi/10.1103/PhysRevResearch.2.033429}
}

@article{
    Hermann_2020_PauliNet,
    title={Deep-neural-network solution of the electronic {S}chrödinger equation},
    volume={12},
    DOI={10.1038/s41557-020-0544-y},
    number={10},
    journal={Nature Chemistry},
    author={Hermann, Jan and Schätzle, Zeno and Noé, Frank},
    year={2020},
    month={Sep},
    pages={891–897},
}

@misc{Spencer_2020_better,
    title={Better, Faster Fermionic Neural Networks}, 
    author={James S. Spencer and David Pfau and Aleksandar Botev and W. M. C. Foulkes},
    year={2020},
    eprint={2011.07125},
    archivePrefix={arXiv},
    primaryClass={physics.comp-ph},
    url={https://arxiv.org/abs/2011.07125}, 
}

@article{Viteritti_2023_spinformer,
    title = {Transformer Variational Wave Functions for Frustrated Quantum Spin Systems},
    author = {Viteritti, Luciano Loris and Rende, Riccardo and Becca, Federico},
    journal = {Phys. Rev. Lett.},
    volume = {130},
    issue = {23},
    pages = {236401},
    numpages = {6},
    year = {2023},
    month = {Jun},
    publisher = {American Physical Society},
    doi = {10.1103/PhysRevLett.130.236401},
    url = {https://link.aps.org/doi/10.1103/PhysRevLett.130.236401}
}

@inproceedings{VonGlehn_2023_PsiFormer,
    title={A Self-Attention Ansatz for Ab-initio Quantum Chemistry},
    author={Ingrid von Glehn and James S Spencer and David Pfau},
    booktitle={The Eleventh International Conference on Learning Representations },
    year={2023},
    url={https://openreview.net/forum?id=xveTeHVlF7j}
}

@article{Hermann_2023_review,
    title={Ab initio quantum chemistry with neural-network wavefunctions},
    volume={7},
    DOI={10.1038/s41570-023-00516-8},
    number={10},
    journal={Nature Reviews Chemistry},
    author={Hermann, Jan and Spencer, James and Choo, Kenny and Mezzacapo, Antonio and Foulkes, W. M. and Pfau, David and Carleo, Giuseppe and Noé, Frank},
    year={2023},
    month={Aug},
    pages={692–709},
}

@article{Lange_2024_review,
    doi = {10.1088/2058-9565/ad7168},
    url = {https://doi.org/10.1088/2058-9565/ad7168},
    year = {2024},
    month = {sep},
    publisher = {IOP Publishing},
    volume = {9},
    number = {4},
    pages = {040501},
    author = {Lange, Hannah and Van de Walle, Anka and Abedinnia, Atiye and Bohrdt, Annabelle},
    title = {From architectures to applications: a review of neural quantum states},
    journal = {Quantum Science and Technology},
}

@article{
    Yoshioka_2021_band,
    title={Solving quasiparticle band Spectra of real solids using neural-network quantum states},
    volume={4},
    DOI={10.1038/s42005-021-00609-0},
    number={1},
    journal={Communications Physics},
    author={Yoshioka, Nobuyuki and Mizukami, Wataru and Nori, Franco},
    year={2021},
    month={May},
}

@article{Lange_2024_quasiparticles,
    title={Neural network approach to quasiparticle dispersions in doped antiferromagnets},
    volume={7},
    DOI={10.1038/s42005-024-01678-7},
    number={1},
    journal={Communications Physics},
    author={Lange, Hannah and Döschl, Fabian and Carrasquilla, Juan and Bohrdt, Annabelle},
    year={2024},
    month={Jun},
}

@article{Fu_2024_qiankun,
    title={Transformer-based neural-network quantum state method for electronic band structures of Real Solids},
    volume={20},
    DOI={10.1021/acs.jctc.4c00567},
    number={14},
    journal={Journal of Chemical Theory and Computation},
    author={Fu, Lizhong and Wu, Yangjun and Shang, Honghui and Yang, Jinlong},
    year={2024},
    month={Jul},
    pages={6218–6226},
}

@misc{Zhang_2025_NTB,
    title={Neural Transformer Backflow for Solving Momentum-Resolved Ground States of Strongly Correlated Materials}, 
    author={Lixing Zhang and Di Luo},
    year={2025},
    eprint={2509.09275},
    archivePrefix={arXiv},
    primaryClass={cond-mat.str-el},
    url={https://arxiv.org/abs/2509.09275}, 
}

@article{Fu_2024_variance,
    doi = {10.1088/2632-2153/ad1f75},
    url = {https://doi.org/10.1088/2632-2153/ad1f75},
    year = {2024},
    month = {jan},
    publisher = {IOP Publishing},
    volume = {5},
    number = {1},
    pages = {015016},
    author = {Fu, Weizhong and Ren, Weiluo and Chen, Ji},
    title = {Variance extrapolation method for neural-network variational {M}onte {C}arlo},
    journal = {Machine Learning: Science and Technology},
}

@article{Amari_1998_natural,
    author = {Amari, Shun-ichi},
    title = {Natural Gradient Works Efficiently in Learning},
    journal = {Neural Computation},
    volume = {10},
    number = {2},
    pages = {251-276},
    year = {1998},
    month = {02},
    issn = {0899-7667},
    doi = {10.1162/089976698300017746},
    url = {https://doi.org/10.1162/089976698300017746},
}

@article{Sorella_1998_green,
    title = {Green Function {M}onte {C}arlo with Stochastic Reconfiguration},
    author = {Sorella, Sandro},
    journal = {Phys. Rev. Lett.},
    volume = {80},
    issue = {20},
    pages = {4558--4561},
    numpages = {0},
    year = {1998},
    month = {May},
    publisher = {American Physical Society},
    doi = {10.1103/PhysRevLett.80.4558},
    url = {https://link.aps.org/doi/10.1103/PhysRevLett.80.4558}
}

@article{Sorella_2001_lanczos,
    title = {Generalized {L}anczos algorithm for variational quantum {M}onte {C}arlo},
    author = {Sorella, Sandro},
    journal = {Phys. Rev. B},
    volume = {64},
    issue = {2},
    pages = {024512},
    numpages = {16},
    year = {2001},
    month = {Jun},
    publisher = {American Physical Society},
    doi = {10.1103/PhysRevB.64.024512},
    url = {https://link.aps.org/doi/10.1103/PhysRevB.64.024512}
}

@article{Li_2024_folx,
    title={A computational framework for neural network-based variational {Monte Carlo} with forward laplacian},
    volume={6},
    DOI={10.1038/s42256-024-00794-x},
    number={2},
    journal={Nature Machine Intelligence},
    author={Li, Ruichen and Ye, Haotian and Jiang, Du and Wen, Xuelan and Wang, Chuwei and Li, Zhe and Li, Xiang and He, Di and Chen, Ji and Ren, Weiluo and et al.},
    year={2024},
    month={Feb},
    pages={209–219},
}

@article{Chen_2024_miNSR,
    author={Chen, Ao and Heyl, Markus},
    title={Empowering deep neural quantum states through efficient optimization},
    journal={Nature Physics},
    year={2024},
    month={Sep},
    day={01},
    volume={20},
    number={9},
    pages={1476-1481},
    issn={1745-2481},
    doi={10.1038/s41567-024-02566-1},
    url={https://doi.org/10.1038/s41567-024-02566-1}
}

@article{Rende_2024_simple,
    author={Rende, Riccardo and Viteritti, Luciano Loris and Bardone, Lorenzo and Becca, Federico and Goldt, Sebastian},
    title={A simple linear algebra identity to optimize large-scale neural network quantum states},
    journal={Communications Physics},
    year={2024},
    month={Aug},
    day={02},
    volume={7},
    number={1},
    pages={260},
    issn={2399-3650},
    doi={10.1038/s42005-024-01732-4},
    url={https://doi.org/10.1038/s42005-024-01732-4}
}

@book{Goodfellow_2016_book,
    title={Deep Learning},
    author={Ian Goodfellow and Yoshua Bengio and Aaron Courville},
    publisher={MIT Press},
    note={\url{http://www.deeplearningbook.org}},
    year={2016}
}

@article{
    Westerhout_2020_generalize,
    title={Generalization properties of neural network approximations to frustrated magnet ground states},
    volume={11},
    DOI={10.1038/s41467-020-15402-w},
    number={1},
    eid = {1593},
    journal={Nature Communications},
    author={Westerhout, Tom and Astrakhantsev, Nikita and Tikhonov, Konstantin S. and Katsnelson, Mikhail I. and Bagrov, Andrey A.},
    year={2020},
    month={Mar},
}

@article{Zhang_2023_spurious,
    title = {Understanding and eliminating spurious modes in variational {Monte Carlo} using collective variables},
    author = {Zhang, Huan and Webber, Robert J. and Lindsey, Michael and Berkelbach, Timothy C. and Weare, Jonathan},
    journal = {Phys. Rev. Res.},
    volume = {5},
    issue = {2},
    pages = {023101},
    numpages = {12},
    year = {2023},
    month = {May},
    publisher = {American Physical Society},
    doi = {10.1103/PhysRevResearch.5.023101},
    url = {https://link.aps.org/doi/10.1103/PhysRevResearch.5.023101}
}

@book{Efron_1994_jackknife,
    place={Boca Raton; London; New York; Washington,D.C},
    title={An introduction to the bootstrap},
    publisher={Chapman et Hall/CRC},
    author={Efron, Bradley and Tibshirani, Robert},
    year={1994}
}

@article{Cassella_2023_wigner,
    title = {Discovering Quantum Phase Transitions with Fermionic Neural Networks},
    author = {Cassella, Gino and Sutterud, Halvard and Azadi, Sam and Drummond, N. D. and Pfau, David and Spencer, James S. and Foulkes, W. M. C.},
    journal = {Phys. Rev. Lett.},
    volume = {130},
    issue = {3},
    pages = {036401},
    numpages = {6},
    year = {2023},
    month = {Jan},
    publisher = {American Physical Society},
    doi = {10.1103/PhysRevLett.130.036401},
    url = {https://link.aps.org/doi/10.1103/PhysRevLett.130.036401}
}

@article{Kato_1957_original,
    author = {Kato, Tosio},
    title = {On the eigenfunctions of many-particle systems in quantum mechanics},
    journal = {Communications on Pure and Applied Mathematics},
    volume = {10},
    number = {2},
    pages = {151-177},
    doi = {https://doi.org/10.1002/cpa.3160100201},
    url = {https://onlinelibrary.wiley.com/doi/abs/10.1002/cpa.3160100201},
    year = {1957}
}

@article{Drummond_2004_jastrow,
    title = {Jastrow correlation factor for atoms, molecules, and solids},
    author = {Drummond, N. D. and Towler, M. D. and Needs, R. J.},
    journal = {Phys. Rev. B},
    volume = {70},
    issue = {23},
    pages = {235119},
    numpages = {11},
    year = {2004},
    month = {Dec},
    publisher = {American Physical Society},
    doi = {10.1103/PhysRevB.70.235119},
    url = {https://link.aps.org/doi/10.1103/PhysRevB.70.235119}
}

@inproceedings{Zhang_2019_RMSNorm,
    author = {Zhang, Biao and Sennrich, Rico},
    booktitle = {Advances in Neural Information Processing Systems},
    editor = {H. Wallach and H. Larochelle and A. Beygelzimer and F. d\textquotesingle Alch\'{e}-Buc and E. Fox and R. Garnett},
    pages = {},
    publisher = {Curran Associates, Inc.},
    title = {Root Mean Square Layer Normalization},
    url = {https://proceedings.neurips.cc/paper_files/paper/2019/file/1e8a19426224ca89e83cef47f1e7f53b-Paper.pdf},
    volume = {32},
    year = {2019}
}

@article{Bahdanau_2014_align,
    title={Neural Machine Translation by Jointly Learning to Align and Translate},
    author={Dzmitry Bahdanau and Kyunghyun Cho and Yoshua Bengio},
    journal={CoRR},
    year={2014},
    volume={abs/1409.0473},
    url={https://api.semanticscholar.org/CorpusID:11212020}
}

@article{Luong_2015_attention,
    title={Effective Approaches to Attention-based Neural Machine Translation},
    author={Thang Luong and Hieu Pham and Christopher D. Manning},
    journal={ArXiv},
    year={2015},
    volume={abs/1508.04025},
    url={https://api.semanticscholar.org/CorpusID:1998416}
}

@inproceedings{Vaswani_2017_transformer,
    author = {Vaswani, Ashish and Shazeer, Noam and Parmar, Niki and Uszkoreit, Jakob and Jones, Llion and Gomez, Aidan N and Kaiser, Lukasz and Polosukhin, Illia},
    booktitle = {Advances in Neural Information Processing Systems},
    editor = {I. Guyon and U. Von Luxburg and S. Bengio and H. Wallach and R. Fergus and S. Vishwanathan and R. Garnett},
    pages = {},
    publisher = {Curran Associates, Inc.},
    title = {Attention is All you Need},
    url ={https://proceedings.neurips.cc/paper_files/paper/2017/file/3f5ee243547dee91fbd053c1c4a845aa-Paper.pdf},
    volume = {30},
    year = {2017}
}

@InProceedings{Xiong_2020_prenorm,
    title = {On Layer Normalization in the {T}ransformer Architecture},
    author = {Xiong, Ruibin and Yang, Yunchang and He, Di and Zheng, Kai and Zheng, Shuxin and Xing, Chen and Zhang, Huishuai and Lan, Yanyan and Wang, Liwei and Liu, Tieyan},
    booktitle = {Proceedings of the 37th International Conference on Machine Learning},
    pages = {10524--10533},
    year = {2020},
    editor = {III, Hal Daumé and Singh, Aarti},
    volume = {119},
    series = {Proceedings of Machine Learning Research},
    month = {13--18 Jul},
    publisher = {PMLR},
    url = {https://proceedings.mlr.press/v119/xiong20b.html},
}

@misc{Hendrycks_2016_GELU,
    title={Gaussian Error Linear Units {(GELUs)}}, 
    author={Dan Hendrycks and Kevin Gimpel},
    year={2023},
    eprint={1606.08415},
    archivePrefix={arXiv},
    primaryClass={cs.LG},
    url={https://arxiv.org/abs/1606.08415}, 
}

@inproceedings{Schutt_2017_SchNet,
    author = {Sch\"{u}tt, K. T. and Kindermans, P.-J. and Sauceda, H. E. and Chmiela, S. and Tkatchenko, A. and M\"{u}ller, K.-R.},
    title = {SchNet: a continuous-filter convolutional neural network for modeling quantum interactions},
    year = {2017},
    isbn = {9781510860964},
    publisher = {Curran Associates Inc.},
    address = {Red Hook, NY, USA},
    booktitle = {Proceedings of the 31st International Conference on Neural Information Processing Systems},
    pages = {992–1002},
    numpages = {11},
    location = {Long Beach, California, USA},
    series = {NIPS'17},
    url = {https://proceedings.neurips.cc/paper_files/paper/2017/file/303ed4c69846ab36c2904d3ba8573050-Paper.pdf},
}

@inproceedings{Gerard_2022_gold,
    title={Gold-standard solutions to the {Schr\"odinger} equation using deep learning: How much physics do we need?},
    author={Leon Gerard and Michael Scherbela and Philipp Marquetand and Philipp Grohs},
    booktitle={Advances in Neural Information Processing Systems},
    editor={Alice H. Oh and Alekh Agarwal and Danielle Belgrave and Kyunghyun Cho},
    year={2022},
    url={https://openreview.net/forum?id=nX-gReQ0OT}
}

@article{Pescia_2024_MPNQS,
    title = {Message-passing neural quantum states for the homogeneous electron gas},
    author = {Pescia, Gabriel and Nys, Jannes and Kim, Jane and Lovato, Alessandro and Carleo, Giuseppe},
    journal = {Phys. Rev. B},
    volume = {110},
    issue = {3},
    pages = {035108},
    numpages = {11},
    year = {2024},
    month = {Jul},
    publisher = {American Physical Society},
    doi = {10.1103/PhysRevB.110.035108},
    url = {https://link.aps.org/doi/10.1103/PhysRevB.110.035108}
}

@article{Geier_2025_moire,
    title = {Self-attention neural network for solving correlated electron problems in solids},
    author = {Geier, Max and Nazaryan, Khachatur and Zaklama, Timothy and Fu, Liang},
    journal = {Phys. Rev. B},
    volume = {112},
    issue = {4},
    pages = {045119},
    numpages = {16},
    year = {2025},
    month = {Jul},
    publisher = {American Physical Society},
    doi = {10.1103/qxc3-bkc7},
    url = {https://link.aps.org/doi/10.1103/qxc3-bkc7}
}

@misc{Dai_2026_barrier,
    title={Essentially No Energy Barrier Between Independent Fermionic Neural Quantum State Minima}, 
    author={David D. Dai and Marin Soljačić},
    year={2026},
    eprint={2601.06939},
    archivePrefix={arXiv},
    primaryClass={cond-mat.dis-nn},
    url={https://arxiv.org/abs/2601.06939}, 
}

@article{Luo_2026_GemiNet,
    title = {Pairing-based graph neural network for simulating quantum materials},
    author = {Luo, Di and Dai, David D. and Fu, Liang},
    journal = {Phys. Rev. B},
    volume = {113},
    issue = {16},
    pages = {165107},
    numpages = {7},
    year = {2026},
    month = {Apr},
    publisher = {American Physical Society},
    doi = {10.1103/5fp1-y42d},
    url = {https://link.aps.org/doi/10.1103/5fp1-y42d}
}

@misc{Sutton_2019_bitter,
  author = {Sutton, Richard S.},
  title = {The Bitter Lesson},
  year = {2019},
  howpublished = {\url{http://www.incompleteideas.net/IncIdeas/BitterLesson.html}},
  note = {Accessed: 2026-06-09}
}

@article{DePalo_2004_charged,
    title = {{Monte Carlo} simulations of two-dimensional charged bosons},
    author = {De Palo, S. and Conti, S. and Moroni, S.},
    journal = {Phys. Rev. B},
    volume = {69},
    issue = {3},
    pages = {035109},
    numpages = {6},
    year = {2004},
    month = {Jan},
    publisher = {American Physical Society},
    doi = {10.1103/PhysRevB.69.035109},
    url = {https://link.aps.org/doi/10.1103/PhysRevB.69.035109}
}

@article{Clark_2009_charged,
    title = {Hexatic and Mesoscopic Phases in a {2D} Quantum Coulomb System},
    author = {Clark, Bryan K. and Casula, Michele and Ceperley, D. M.},
    journal = {Phys. Rev. Lett.},
    volume = {103},
    issue = {5},
    pages = {055701},
    numpages = {4},
    year = {2009},
    month = {Jul},
    publisher = {American Physical Society},
    doi = {10.1103/PhysRevLett.103.055701},
    url = {https://link.aps.org/doi/10.1103/PhysRevLett.103.055701}
}

@article{Boninsegni_2026_plasma,
    title = {Bose one-component plasma in two dimensions: A {Monte Carlo} study},
    author = {Boninsegni, Massimo},
    journal = {Phys. Rev. B},
    volume = {113},
    issue = {9},
    pages = {094508},
    numpages = {6},
    year = {2026},
    month = {Mar},
    publisher = {American Physical Society},
    doi = {10.1103/9rpx-wdjs},
    url = {https://link.aps.org/doi/10.1103/9rpx-wdjs}
}

@article{Nosanow_1962_hartree,
    title = {Hartree Calculations for the Ground State of Solid {He} and other Noble Gas Crystals},
    author = {Nosanow, Lewis H. and Shaw, Gordon L.},
    journal = {Phys. Rev.},
    volume = {128},
    issue = {2},
    pages = {546--550},
    numpages = {0},
    year = {1962},
    month = {Oct},
    publisher = {American Physical Society},
    doi = {10.1103/PhysRev.128.546},
    url = {https://link.aps.org/doi/10.1103/PhysRev.128.546}
}

@article{Magro_1994_log,
    title = {Ground-State Properties of the Two-Dimensional Bose Coulomb Liquid},
    author = {Magro, W. R. and Ceperley, D. M.},
    journal = {Phys. Rev. Lett.},
    volume = {73},
    issue = {6},
    pages = {826--829},
    numpages = {0},
    year = {1994},
    month = {Aug},
    publisher = {American Physical Society},
    doi = {10.1103/PhysRevLett.73.826},
    url = {https://link.aps.org/doi/10.1103/PhysRevLett.73.826}
}

\end{document}